\documentclass{aa}  

\usepackage{graphicx}
\usepackage{txfonts}
\usepackage{xcolor}

\usepackage{mathtools}
\usepackage{empheq}
\usepackage{hyperref}

\usepackage{placeins}
\usepackage{multirow}

\def\br{\mbox{\boldmath$r$}}
\def\be{\mbox{\boldmath$e$}}
\def\bu{\mbox{\boldmath$u$}}
\def\bxi{\boldsymbol \xi}
\def\ii{\textrm{i}}
\def\vdop{\mathrm{v}}

\def\Vref{V_{\rm dop}\,}

\def\Ihmi{hmi.Ic\_45s}
\def\Vhmi{hmi.V\_45s}

\defcitealias{Kostogryz2021}{K+21}

\begin{document}

\title{Modeling HMI observables for the study of solar oscillations}

 \titlerunning{Modeling HMI observables for the study of solar oscillations} 
   \author{ D.~Fournier\inst{1, \thanks{Corresponding authors: fournier@mps.mpg.de,\\ kostogryz@mps.mpg.de, gizon@mps.mpg.de}}
   \and
  N.~Kostogryz\inst{1,^\star}
          \and
          L. Gizon\inst{1,2,^\star}
           \and
          J. Schou\inst{1}
          \and
          V. Witzke\inst{3}
          \and 
          A. I. Shapiro\inst{1,3}
          \and
          I. Mili\'c\inst{4,5,6} }

   \institute{\inst{1} Max-Planck-Institut f\"ur Sonnensystemforschung, Justus-von-Liebig-Weg 3, 37077 G\"ottingen, Germany\\
   \inst{2} Institut f\"ur Astrophysik und Geophysik, Georg-August-Universit\"at G\"ottingen, Friedrich-Hund-Platz 1, 37077 G\"ottingen, Germany\\
   \inst{3} University of Graz, Institute of Physics, Universitätsplatz 5, 8010 Graz, Austria \\
   \inst{4} Institute for Solar Physics (KIS), Schöneck Str. 6, 79111 Freiburg, Germany\\
   \inst{5} Astronomical Observatory, Volgina 7, 11060 Belgrade, Serbia \\
   \inst{6} Department of Astronomy, Faculty of Mathematics, University of Belgrade, Studentski Trg 16-20, 11000, Belgrade, Serbia}
             
\date{Received $\langle$date$\rangle$ / Accepted $\langle$date$\rangle$}

 
  \abstract
   {Helioseismology aims to infer the properties of the solar interior by analyzing observations of acoustic oscillations. The interpretation of  the helioseismic data is however complicated by the non-trivial relationship between helioseismic observables and the physical perturbations associated with  acoustic modes, as well as by various  instrumental effects. 
   }
   {We aim to improve our understanding of the signature of acoustic modes  measured in  the Helioseismic and Magnetic Imager (HMI) continuum intensity and Doppler velocity observables by accounting for radiative transfer, solar background rotation, and spacecraft velocity.}
   {We start with a background model atmosphere that accurately reproduces solar limb darkening and the Fe I 6173$~\AA$ spectral line profile. We employ first-order perturbation theory to model the effect of acoustic oscillations on inferred intensity and velocity. By solving the radiative transfer equation in the atmosphere, we synthesize the spectral line, convolve it with the six HMI spectral windows, and deduce continuum intensity (\Ihmi) and Doppler velocity (\Vhmi) according to the HMI algorithm.}
   {We analytically derive the relationship between mode displacement in the atmosphere and the HMI observables, and show that both  intensity and velocity  deviate significantly from simple approximations. Specifically, the continuum intensity does not simply reflect the true continuum value, while the line-of-sight velocity does not correspond to a straightforward projection of the velocity at a fixed height in the atmosphere. Our results indicate that these deviations are substantial, with amplitudes of approximately 10\% and phase shifts of around $10^\circ$ across the detector for both observables. Moreover, these effects are highly dependent on the acoustic mode under consideration and the position on the solar disk. To achieve accurate modeling of the observables, it is important to account for the impact of radiative transfer on oscillation velocities and perturbations in atmospheric thermodynamic quantities, which influence the line profile.   
   }
   {The combination of these effects leads to non-trivial systematic errors (in amplitude and phase) across the detector that must be taken into account to understand the observables. This framework can be used to study mode visibility across the solar disk and  for asteroseismology applications. }

   \keywords{Radiative transfer - Sun: helioseismology - Sun: atmosphere - Sun: oscillations - Methods: numerical}

   \maketitle
%


\section{Introduction}

Key aspects of helioseismology rely on a good understanding of the helioseismic observables.  
The relationship between  intensity fluctuations and low-degree p modes has been extensively studied \citep[see for example][and references therein]{Toutain1993}. In \citet{Kostogryz2021}  \citepalias[hereafter][]{Kostogryz2021}, we extended this analysis to high-degree modes, incorporating radiative transfer effects and accounting for horizontal wave motions.
Building on \citetalias{Kostogryz2021}, we advance this line of research in several significant ways. 
Firstly, we employ a more realistic background  atmospheric model computed using the Merged Parallelized Simplified ATLAS code \citep[\texttt{MPS-ATLAS},][]{2021A&A...653A..65W}. This model atmosphere accurately reproduces the observed limb darkening and, consequently, accounts for the spectral line formation height \citep{Kostogryz2022}. Secondly, we model the full spectral line, rather than just the continuum intensity, and compute HMI observables from the model filtergrams using the algorithm described by  \citet{Couvidat2012} and referred to as the HMI LoS Pipeline. Specifically, we focus on two observables: the line-of-sight velocity (\Vhmi) and the continuum intensity (\Ihmi). It is worth noting that the continuum intensity computed by the HMI algorithm differs from the theoretical continuum intensity as derived for example by \citetalias{Kostogryz2021}. 

The observables \Vhmi\ and \Ihmi\ are susceptible to various systematic errors due to instrumental limitations and potential misinterpretation due to incomplete physical modeling. While some corrections are applied in the LoS Pipeline \citep{Couvidat2016}, residual systematic effects persist and are addressed differently depending on the helioseismic method. For instance,  \citet{Bogart2015} and \citet{Liang2018} 
have investigated systematic errors in ring-diagram analysis and time-distance helioseismology, respectively. In this work, we aim to elucidate the impact of radiative transfer on the observed oscillations. Furthermore, we incorporate the effects of solar differential rotation and spacecraft velocity, the latter being identified as the most significant remaining systematic error in the observations \citet{Couvidat2016}. 

The remainder of this paper is organized as follows. In Sect.~\ref{sect:background}, we present the spectral line obtained after applying radiative transfer in our background atmosphere. We then summarize the main steps of the HMI algorithm to compute observables from the intensity along the spectral line in Sect.~\ref{sect:HMI}. In Sect.~\ref{sect:firstOrder}, we derive analytically the first-order expressions for the observables when the background model is perturbed by oscillations. These expressions are then evaluated numerically to obtain HMI observables: line-of-sight velocity (\Vhmi) in Sect.~\ref{sect:velocity} and continuum intensity (\Ihmi) in Sect.~\ref{sect:intensity}. Finally, we present our conclusions and outline some possible extensions of this work in Sect.~\ref{sect:discussion}.

\section{Spectral synthesis in the background model} \label{sect:background}

The \(\mathrm{Fe \, I}\) 6173 \AA{} line is observed by the HMI and {Polarimetric and Helioseismic Imager} \citep[PHI,][] {Solanki2020} space-based instruments, and often used for photospheric diagnostics in ground-based observations \citep{Cavallini2006, Scharmer2007}. This line was also chosen for  the {Photospheric Magnetic Field Imager} \citep[PMI;][]{Staub2020} to be flown onboard ESA’s upcoming Vigil space mission.  It is
a diagnostically important photospheric spectral line used because of its large magnetic sensitivity (Land\'{e} factor of 2.5) and lack of (strong) blends that enables robust velocity inference. 

We calculate the emergent intensity $I_\lambda$ at wavelength $\lambda$ (around the line central wavelength) for each point on the visible hemisphere using the following form of the formal solution of the radiative transfer equation:
\begin{equation}
    I_\lambda(\mu) = \int_0^\infty S_\lambda(\tau_\lambda) \textrm{e}^{-\tau_\lambda / \mu} \frac{\mathrm{d}\tau_\lambda}{\mu}, \label{eq:intensity}
\end{equation}
where $\mu$ is the cosine of the angle between the line-of-sight vector and the local normal to the surface, with $\mu=1$ corresponding to disk center and $\mu=0$ to the limb. 
The source function $S_\lambda$ is chosen as a Planck function by assuming local thermodynamic equilibrium. The optical depth $\tau_\lambda$ is obtained from the opacity $\alpha_\lambda$ 
\begin{equation}
    \tau_\lambda(s; \mu) = - \int_s^\infty \alpha_\lambda(s'; \mu) \mathrm{d}s', \label{eq:opticalDepth}
\end{equation}
where $s$ is the geometric height. Numerically, the integration in Eq.~\eqref{eq:intensity} needs to be performed only on the optically thin layers (we use layers where the optical depth in the continuum is between $10^{-8}$ and $60$).

\subsection{Background model} \label{sect:background_model}

A background model (temperature, pressure, density) that accurately reflects the conditions in the solar atmosphere is crucial for realistic modeling of the spectral line. We use a plane-parallel atmosphere, which is a reasonable approximation to compute radiative transfer for the Sun (except very close to the limb), as the atmosphere is thin compared to the solar radius \citep[see, for example,][for a comparison of intensity in plane-parallel and spherical geometries]{Toutain1999}.

\begin{figure}
    \centering
    \includegraphics[width=\linewidth]{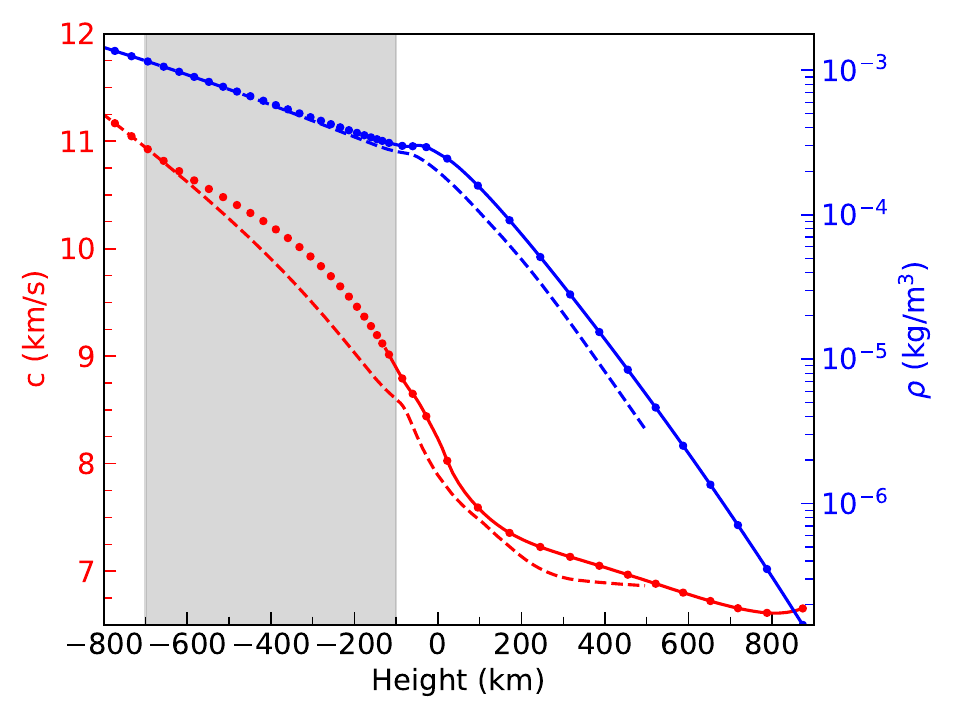}
    \caption{Background sound speed (red) and density (blue) from \texttt{Model~S} (dashed), \texttt{MPS-ATLAS} (solid), and the patched model used in this study (dots). The gray region represents the domain where the patching between \texttt{Model~S} and \texttt{MPS-ATLAS} is done. The zero level of height in this context is defined at one solar radius.}
    \label{fig:background_model}
\end{figure}

We opt for a solar atmosphere model from a grid of stellar models by \citet{Kostogryz2022}, computed with the \texttt{MPS-ATLAS} code, adopting chemical abundances from \citet{2009ARAA..47..481A}. This model includes convection in the lower atmospheric layers, using the mixing-length approximation, and accounts for overshoot from the convective zone into the atmosphere, extending up to one pressure scale height. This background model has been extensively tested against solar measurements \citep{2021A&A...653A..65W, Kostogryz2022}, showing very good agreement with solar limb darkening observations. This implies that the model allows an accurate treatment of the formation height dependence on disk position and justifies its choice for the present study. 

To model oscillations up to the photosphere (where the observed signal is formed) we use a global eigenvalue solver which requires a model for the solar interior. We use the standard solar \texttt{Model~S} \citep{modelS} from the solar center to 700~km below the surface. We smoothly patch it to the atmospheric model from \citet{Kostogryz2022}. A smooth transition between these two models is done close to the surface as shown in Fig.~\ref{fig:background_model} for density and sound speed. We note that any type of perturbation can be used in the setup presented here. We could have also use outputs from numerical simulations or eigenfunctions computed in a Cartesian box using the plane-parallel approximation.

\subsection{Opacity of the \(\mathrm{Fe \, I}\) 6173 \AA{} line and continuum}

\begin{figure*}[!htb]
    \centering
    \begin{tabular}{cc}
    \includegraphics[width=0.5\linewidth]{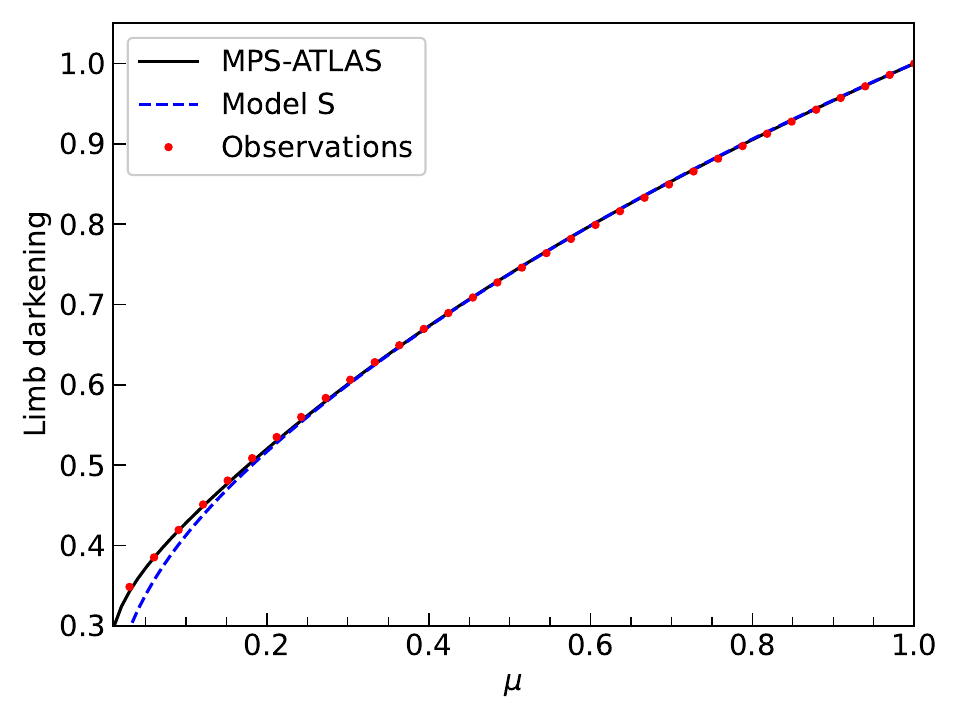} & \includegraphics[width=0.5\linewidth]{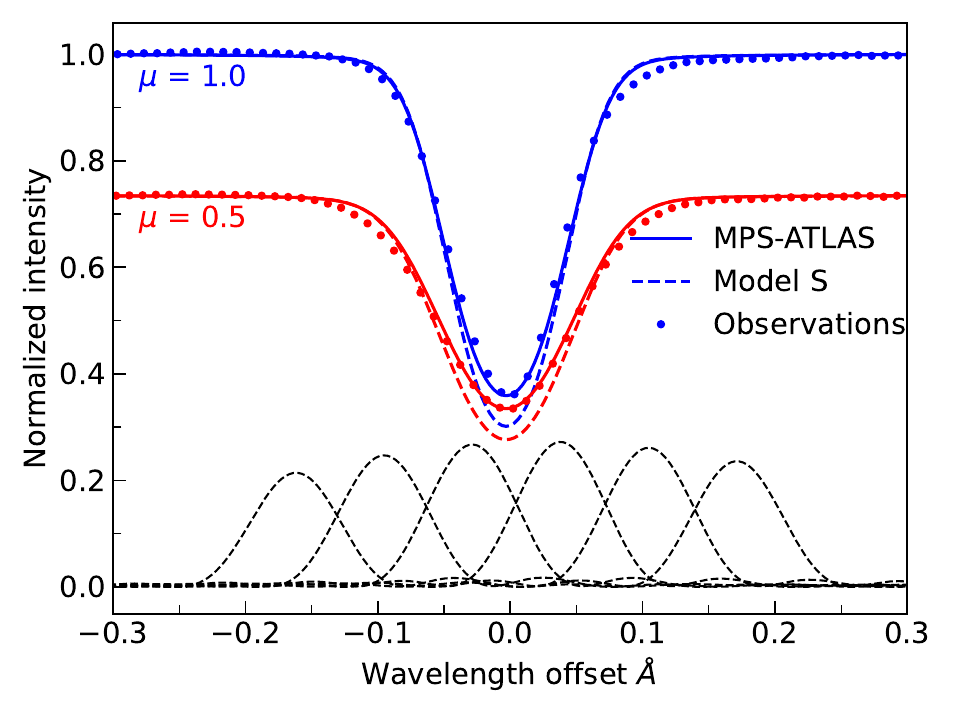}
    \end{tabular}
    \caption{Background intensity computed from the \texttt{MPS-ATLAS}  solar model atmosphere (solid), the standard solar \texttt{Model~S} (dashed), and from the observations (dots). Left: limb darkening ($I_c(\mu) / I_c(\mu=1)$) in the continuum around HMI line compared with the observations from \citet{Neckel1994}. Right: synthesized $\mathrm{Fe I}$ spectral line background intensity ($\tilde{I}(\lambda,\mu) / I_c(\mu=1)$) compared with the observations from IAG spectral atlas \citep{Ellwarth2023} at $\mu = 1$ (blue) and $\mu=0.5$ (red). The wavelength offset is with respect to the center of the \textsc{HMI} line $\lambda_{\rm HMI} = 6173.33$~\AA. The black dashed curves are the six \textsc{HMI} filtergrams.}
    \label{fig:I0}
\end{figure*}

We compute opacity $\alpha_\lambda$ as a sum of the \(\mathrm{Fe \, I}\) line opacity and the continuum opacity employing the high-resolution mode of the \texttt{MPS-ATLAS} code. For the line opacity computation, we select the \(\mathrm{Fe \, I}\) line at \(\lambda = 6175.04 \AA{}\) from the list of atomic lines provided by Vienna Atomic Line Database \citep[VALD;][]{Piskunov1995, Kupka1999, Ryabchikova2015} which correspond to \(\lambda = 6173.33 \AA{}\) after conversion from vacuum to air wavelength. The shape of the spectral line opacity is approximated by a Voigt profile. In addition, we account for the Doppler broadening using a depth-independent micro-turbulent velocity of 1~km/s.

To compute the continuum opacities, we include free-free and bound-free transitions in $\mathrm{H^{-}}$, $\mathrm{H}$ , $\mathrm{He}$, $\mathrm{He}^-$, $\mathrm{C}$, $\mathrm{N}$, $\mathrm{O}$, $\mathrm{Ne}$, $\mathrm{Mg}$, $\mathrm{Al}$, $\mathrm{Si}$, $\mathrm{Ca}$, $\mathrm{Fe}$, the molecules $\mathrm{CH}$, $\mathrm{OH}$, and $\mathrm{NH}$, and their ions for the absorption continuum opacity calculation. For the scattering contribution, we consider electron scattering and Rayleigh scattering on $\mathrm{H I}$, $\mathrm{He I}$, and $\mathrm{H_2}$.

\subsection{Spectral line broadening}

We synthesize the \(\mathrm{Fe \, I}\) spectral line in a 1D geometry by solving Eq.~\ref{eq:intensity} with constant micro-turbulent velocity, resulting in a symmetric (unperturbed) spectral line profile.
This approach neglects the upflow and downflow motions caused by granulation, which contribute to line broadening and introduce line asymmetries. Accurately modeling these effects would require synthesizing spectra using three-dimensional hydrodynamic simulation or using a depth-dependent micro-turbulent velocity, which is beyond the scope of this paper.
However, to account for line broadening due to granulation, we compute the emergent intensity by convolving the synthesized intensity profile with a broadening kernel \citep{Gray2021}:
\begin{equation}
    \tilde{I}(\lambda,\mu) = \left[ I(\mu) \ast G_{\rm macro}(\mu) \right](\lambda),
\end{equation}
where $\ast$ denotes a convolution with respect to wavelength, $G_{\rm macro}$ is a Gaussian kernel of standard deviation $\sigma_{\rm macro}$. To account for the effect that granulation broadens spectral lines differently from the center to the limb, we adopt the standard deviation proposed by \cite{Takeda2019} that depends on the position on the disk:
\begin{equation}
    \sigma_{\rm macro} = \sigma_0 + \sigma_1 \sqrt{ 1-\mu^2}.
\end{equation}

To achieve the correct line broadening, we tune the parameters $\sigma_0$ and $\sigma_1$ to match spectral line observations at different $\mu$-positions from \citet{Ellwarth2023} and obtain $\sigma_0 = 0.95$~km/s and $\sigma_1 = 0.5$~km/s. 

\subsection{Center-to-limb variation of background intensity}

In Figure~\ref{fig:I0}, we compare observed and modeled center-to-limb intensity variations in the continuum (left panel) and spectral line (right panel), using the \texttt{MPS-ATLAS} background model. We also compare with \texttt{Model S} as it was used in our previous study to compute intensity \cite{Kostogryz2021} and is often used in helioseismology. Here, the background intensity refers to the emergent intensity computed in an unperturbed atmosphere, serving as the reference for subsequent analysis. The limb darkening observation was conducted by \citet{Neckel1994} in quiet regions of the Sun. They fitted a fifth-order polynomial in $\mu$ in the continuum at various wavelengths, and we select the values at 6110~\AA{} for comparison.
Both models align well with the limb-darkening observational data for $\mu \geq 0.2$. For lower values of $\mu$, the \texttt{MPS-ATLAS} background model reproduces more accurately the observations. The close alignment in center-to-limb variations in emergent intensity between observations and models indicates that the background model reliably represents the atmospheric structure where continuum forms. 

We also compare the spectral line profiles between our models and the observations  from \citet{Ellwarth2023} at two positions on the disk.
The intensity computed with the \texttt{MPS-ATLAS} model reproduces well the observed spectral line depth and width, while the spectral line depth obtained from \texttt{Model S} is too deep. It shows that the atmospheric layers where the line core forms are not realistic for \texttt{Model S}. Since our objective is to compute Doppler velocities, it is crucial to accurately reproduce the spectral line around the instrument’s central wavelength and therefore use the \texttt{MPS-ATLAS} background model in the remainder of the paper.

\section{HMI algorithm for continuum intensity and velocity} \label{sect:HMI}

The observed intensities are acquired at a small number of filtergrams $N$ (six for HMI), corresponding to the convolution of the wavelength-dependent intensities with filters $F_j$ centered around specific wavelengths $\lambda_j$
\begin{equation}
    \mathcal{I}_j(\mu) = \int_{-\infty}^\infty F_j(\lambda) \tilde{I}(\lambda,\mu) \mathrm{d}\lambda. \label{eq:I_filter}
\end{equation}

In Figure ~\ref{fig:I0}, we show the spectral shapes of the HMI filters at disk center. Here, two filters are primarily located in the continuum, two in the wings of the line, and two closer to the line core. Since the filters are distributed near the central wavelength of the line, the integral in Eq.~\eqref{eq:I_filter} can be evaluated over a narrow wavelength range. We consider only one spectral line and compute Eq.~\eqref{eq:I_filter} as
\begin{equation}
    \mathcal{I}_j(\mu) = \int_{\lambda_{\rm HMI}-0.3}^{\lambda_{\rm HMI}+0.3} F_j(\lambda) \tilde{I}(\lambda,\mu) \mathrm{d}\lambda + I_c(\mu) \int_{|\lambda-\lambda_{\rm HMI}| > 0.3} F_j(\lambda) \mathrm{d}\lambda,
\end{equation}
where $\lambda_{\rm HMI}$ is expressed in \AA. The second integral takes into account that the filters are non-zero in a broader range of frequencies. We assumed that $I_\lambda = I_c$ for $|\lambda-\lambda_{\rm HMI}| > 0.3$~\AA{} and neglected the (weak) blends from other lines for simplicity. 

From the six intensities, the main quantities to compute continuum intensities and velocities using the HMI algorithm are the Fourier coefficients $a_1$ and $b_1$
\begin{align}
    a_1(\mu) &= \frac{2}{N} \sum_{j=1}^N \mathcal{I}_j(\mu) \cos \left( 2\pi \frac{N - 0.5 - j}{N} \right), \label{eq:a1} \\ 
    b_1(\mu) &= \frac{2}{N} \sum_{j=1}^N \mathcal{I}_j(\mu) \sin \left( 2\pi \frac{N - 0.5 - j}{N} \right). \label{eq:b1}
\end{align}
The second Fourier coefficients $a_2$ and $b_2$ are computed in the same manner by replacing $2\pi$ by $4\pi$ in the $\cos$ and $\sin$.

\subsection{Velocity}

The Doppler velocity is computed as \citep{Couvidat2012}
\begin{equation}
    \vdop(\mu) = \Vref \textrm{atan2} [ b_1(\mu), a_1(\mu) ], \label{eq:velocity}
\end{equation}
where the coefficient $\Vref$ is usually written as
\begin{equation}
    \Vref = \frac{\mathrm{d}v}{\mathrm{d}\lambda} \frac{T}{2\pi}.
\end{equation}
The period of the observation wavelength span is $T=412.8$~m\AA \ for HMI, which gives $\Vref \approx 2.91~\textrm{km}/\textrm{s}$. However, it needs to be calibrated via a look-up table due to the limited number of available wavelengths and the filters \citep{Scherrer1995,Couvidat2012}.
The HMI Doppler velocity (\Vhmi) \ is thus given as
\begin{equation}
    \vdop_{\rm HMI}  = \mathcal{F}_{\rm HMI}(\vdop, \mu), \label{eq:lookup}
\end{equation}
where $\mathcal{F}_{\rm HMI}$ corresponds to the look-up table (see Appendix~\ref{sect:lookup} for more details). For not too strong background velocities (less than 3~km/s), this function is almost the identity so that $\vdop_{\rm HMI} \approx \vdop$ (see Fig.~\ref{fig:lookup_equator}). However, it deviates from the identity for large background velocities and/or small values of $\mu$.

\subsection{Continuum intensity}

As for velocity, continuum intensity (\Ihmi) \ is computed from the Fourier coefficients
\begin{equation}
I_{\rm HMI}(\mu) = \frac{1}{6} \sum_{j=1}^N \left[ \mathcal{I}_j(\mu) + I_d(\mu) \ \textrm{exp} \left( - \frac{(\lambda_j - \lambda_{\rm HMI})^2}{\sigma^2(\mu)} \right) \right], \label{eq:Ihmi}
\end{equation}
where the estimate of the line width is
\begin{equation}
    \sigma(\mu) = \frac{T}{\pi \sqrt{6}} \sqrt{\log \left( \frac{a_1^2(\mu) + b_1^2(\mu)}{a_2^2(\mu)+b_2^2(\mu)} \right)},
\end{equation}
and of the line depth 
\begin{equation}
    I_d(\mu) = \frac{T}{2\sqrt{\pi}\sigma(\mu)} \sqrt{a_1^2(\mu)+b_1^2(\mu)} \exp \left( \frac{\pi^2\sigma^2(\mu)}{T^2} \right).
\end{equation}

\section{First-order perturbations for observables} \label{sect:firstOrder}

Similarly to the approach used by \citetalias{Kostogryz2021} for the theoretical continuum intensity, we now want to evaluate the perturbed intensity and velocity caused by the oscillations. We linearize around a background state which represents a spherically symmetric Sun characterized by its temperature $T_0$, pressure $p_0$, background velocity $\bu_0$ (for example, rotation), and normal vector $\br_0 = r_0 \textbf{e}_r$, at the unperturbed radial coordinate $r_0$. The oscillations perturb the surface
\begin{equation}
\br = \br_0 + \bxi,
\end{equation}
where $\bxi$ is the Lagrangian displacement vector, which is linked to the Eulerian velocity through
\begin{equation}
    \bu = \partial_t \bxi -  \bxi \cdot \nabla \bu_0.
\end{equation} 
The oscillations also modify the thermodynamical quantities  $T = T_0 + \delta T$, $p = p_0 + \delta p$, where the $\delta$ corresponds to Lagrangian perturbations. 
In this paper, we perturb the surface by the oscillations of a single global mode (see Sect.~\ref{sect:velocity}), but the theoretical framework presented below is general. A summary of the different quantities in their background and perturbed state is given in Tab.~\ref{tab:summary_quantities}.

\begin{table*}[!htb]
    \centering
    \caption{Summary of the main quantities defined in the paper. The perturbed HMI velocity $\delta\vdop_{\rm HMI}$ is compared to the approximation $\delta \vdop_{\rm los}$ and the HMI continuum $\delta I_{\rm HMI}$ to the pure continuum $\delta I_c$.}
    \begin{tabular}{lll}
    \hline
\hline
        Quantity & Background &  Perturbations\\
        \hline
        \multirow{2}{*}{Intensity $I_\lambda$} & \multirow{2}{*}{$I_\lambda^0$ Eq.~\eqref{eq:intensity}} & $\delta I_\lambda = \delta I_{\rm geom} + \delta I_{\rm th} + \delta I_{\rm line}$ \\
        &  & Eqs.~\eqref{eq:deltaIXi}, \eqref{eq:pert_vel}, and \eqref{eq:pert_vel} \\
        HMI filtergrams $\mathcal{I}_j$ ($1 \leq j \leq 6$) & $\mathcal{I}_j^0$ (Eq.~\eqref{eq:I_filter} with $I_0^\lambda$) & $\delta \mathcal{I}_j$ (Eq.~\eqref{eq:I_filter} with $\delta I_\lambda$) \\
        Fourier coefficients $a_1$ and $b_1$ & $a_1^0$, $b_1^0$ (Eqs.~\eqref{eq:a1}, \eqref{eq:b1} with $\mathcal{I}_j^0$) & $\delta a_1$, $\delta b_1$ (Eqs.~\eqref{eq:a1}, \eqref{eq:b1} with $\delta\mathcal{I}_j$) \\
        Velocity $\vdop$ (before lookup table) & $\vdop^0$ (Eq.~\eqref{eq:v0}) & $\delta\vdop$ (Eq.~\eqref{eq:deltav}) \\
        \hline
        \multirow{2}{*}{$\vdop_{\rm HMI}$ (corresponding to \Vhmi)}  & \multirow{2}{*}{$\vdop^0_{\rm HMI}$ \eqref{eq:v0_HMI}} & $\delta \vdop_{\rm HMI}$ \eqref{eq:deltav_HMI} \\
         & & $= \delta\vdop_{\rm geom} + \delta\vdop_{\rm th} + \delta\vdop_{\rm line} $ (from $\delta I_{\rm geom}$, $\delta I_{\rm th}$, $\delta I_{\rm line}$) \\
        Line-of-sight flow velocity & $\vdop^0_{\rm los}$ (Eq.~\eqref{eq:ulos} with $\bu=\bu_0$) & $\delta \vdop_{\rm los}$ (approximation at fixed height, Eq.~\eqref{eq:delav_los}) \\
        \hline
        $I_{\rm HMI}$ (corresponding to \Ihmi) & $I_{\rm HMI}^0$ \eqref{eq:Ihmi} with $\mathcal{I}_j^0$ & $\delta I_{\rm HMI}$ \eqref{eq:deltaIc} \\
        Continuum intensity $I_c$ & $I_c^0$ (=$I^0_\lambda$ for $\lambda$ in continuum) & $\delta I_c$ (=$\delta I_\lambda$ for $\lambda$ in continuum) \\
        \hline
    \end{tabular}
    \label{tab:summary_quantities}
\end{table*}

This approach is in line with the idea of response functions used in solar spectropolarimetric diagnostics \citep[e.g.][]{Milkey1975_rf, LL1977_rf, Milic2017_rf}. The idea behind the response functions is to devise a kernel that expresses sensitivity of the emergent (generally, polarized) spectrum to the perturbations of specific physical quantities (temperature, pressure, velocity, magnetic field) in the atmosphere. This kernel is then used to design the observations or to fit the model atmosphere to the observed spectra, i.e., conduct spectropolarimetric inversion. Recently, an interesting application of the response functions, following concepts from helioseismology, has been proposed by \cite{Agrawal2023rf}

\subsection{Perturbed  intensity} \label{sect:firstOrderI}

In the continuum, the perturbed intensity can be written as a sum of purely thermodynamical terms (depending only on $\delta T$ or $\delta p$), and geometrical terms (depending on all the components of the displacement $\boldsymbol{\xi}$) \citepalias[see e.g.][]{Kostogryz2021}. In the line, an additional term corresponding to a Doppler shift appears, and the perturbed intensity can be written as
\begin{equation}
    \delta I_\lambda(\mu) = \delta I_{\rm th}(\lambda,\mu) + \delta I_{\rm geom}(\lambda,\mu) + \delta I_{\rm line}(\lambda,\mu). \label{eq:deltaI}
\end{equation}
This last term is due to the dependency of opacity on the line-of-sight velocity
\begin{equation}
\vdop^0_{\rm los} = \bu \cdot \mathbf{e}_{\rm obs},  \label{eq:ulos}
\end{equation}
where $\mathbf{e}_{\rm obs}$ points toward the observer (see e.g. Eq.~(A.12) in \citetalias{Kostogryz2021} for its expression in spherical coordinates). The variations of the opacity $\alpha(p,T,\vdop_{\rm los})$ are obtained from
\begin{equation}
    \frac{\delta \alpha_\lambda}{\alpha_\lambda^0} = \frac{\partial(\log \alpha_\lambda^0)}{\partial (\log p_0)} \frac{\delta p }{p_0} + \frac{\partial(\log \alpha_\lambda^0)}{\partial (\log T_0)} \frac{\delta T }{T_0} + \frac{\partial(\log \alpha_\lambda^0)}{\partial \vdop^0_{\rm los}} \delta \vdop_{\rm los},
\end{equation}
where the derivative in the last term can be calculated as
\begin{equation}
    \frac{\partial(\log \alpha_\lambda^0)}{\partial \vdop^0_{\rm los}} = \frac{\partial(\log \alpha_\lambda^0)}{\partial \lambda} \frac{\partial \lambda}{\partial \vdop^0_{\rm los}} =  \frac{\partial(\log \alpha_\lambda^0)}{\partial \lambda} \frac{\lambda}{c}.
\end{equation}
The thermodynamical and geometrical terms are derived in \citetalias{Kostogryz2021} and we recall their formulation here for completeness. The geometrical term depends on the displacement and its derivatives and is given by
    \begin{align}
    \delta I_{\rm geom}(\lambda,\mu) = & \frac{1}{\mu_0}\int_{0}^{\infty} r_0 S_\lambda^0 \, e^{-{\tau_\lambda^0}/{\mu_0}} \Biggl\{  
    \frac{\partial }{\partial r_0} \left( \frac{\boldsymbol{\xi} \cdot \be_{\textrm{obs}}}{r_0} \right)
    \nonumber \\ & 
    +  \left( \frac{\tau_\lambda^0}{\mu_0}-1 \right) \left( \mu_0 \frac{\partial}{\partial r_0} \left( \frac{\xi_r}{r_0} \right) - \frac{\nabla \boldsymbol{\xi} \cdot \be_{\mathrm{obs}}}{r_0}  \right) \Biggr\} \frac{\mathrm{d}\tau_\lambda^0}{\mu_0}.
    \label{eq:deltaIXi}
\end{align}
The thermodynamical term reflects the changes in the opacity and source function and can be written as
\begin{equation}
\delta I_{\rm th}(\lambda,\mu) = \int_0^\infty \left[ f_T^\lambda \frac{\delta T}{T_0} + f_P^\lambda \frac{\delta p}{p_0} \right] \, \textrm{e}^{-\tau_\lambda^0/\mu_0} \,  \frac{\mathrm{d}\tau_\lambda^0}{\mu_0}, \label{eq:deltaI_th}
\end{equation}
where the functions $f_T$ and $f_P$ are given by
\begin{align}
    f_T^\lambda(\tau_\lambda^0,\mu_0) &= S_\lambda^0  \frac{ h \nu / k T_0}{1 - \textrm{e}^{- h \nu / k T_0}} +  \frac{\partial \log \alpha^0_\lambda}{\partial \log T_0} \, \Bigl[ S^0_\lambda   - I^0_\lambda \Bigr],
    \\
    f_P^\lambda(\tau_\lambda^0,\mu_0) &=  \frac{\partial \log \alpha^0_\lambda}{\partial \log p_0} \, \Bigl[ S^0_\lambda - I^0_\lambda \Bigr].
\end{align}
The relation from \citetalias{Kostogryz2021} has been integrated by parts to make appear the background intensity $I_\lambda^0$, an expression already obtained by \citet{Zhugzhda1996} (see Appendix~\ref{sect:thermodynamic}). Using Eq.~(21) from \citetalias{Kostogryz2021} that relates the perturbed intensity to the change in opacity, we can obtain the perturbations due to the line-of-sight velocity 
\begin{equation}
    \delta I_{\rm line}(\lambda,\mu) =  \int_0^\infty \frac{\partial \log \alpha^0_\lambda}{\partial \lambda} \, \frac{\lambda}{c} \, \Bigl[ S^0_\lambda - I^0_\lambda \Bigr] \delta\vdop_{\rm los} \, \textrm{e}^{-\tau^0_\lambda/\mu_0} \,  \frac{\mathrm{d}\tau^0_\lambda}{\mu_0}. \label{eq:pert_vel}
\end{equation}
This term shifts the spectral line and is the main contributor to the velocity as we will see in the numerical section (Sect.~\ref{sect:velocity}).

\subsection{Perturbed velocity}

We show in App.~\ref{sect:first_order_vel} that $\vdop = \vdop^0 + \delta \vdop$ where
\begin{align}
\vdop^0 &= \Vref \textrm{atan2} [b_1^0, a_1^0], \label{eq:v0} \\
    \delta \vdop &= \Vref \frac{\sum_{i=1}^N\sum_{j=1}^N \mathcal{I}_i^0 \delta \mathcal{I}_j \sin\left( 2\pi \frac{j-i}{N} \right)}{\sum_{i=1}^N\sum_{j=1}^N \mathcal{I}_i^0  \mathcal{I}_j^0 \cos\left( 2\pi \frac{j-i}{N} \right)} \label{eq:deltav}.
\end{align}
The perturbed filtergrams $\delta\mathcal{I}_j$ such that $\mathcal{I}_j = \mathcal{I}_j^0 + \delta \mathcal{I}_j$ are computed from Eq.~\eqref{eq:I_filter} using the perturbed intensities derived in the previous section. 

The look-up table needs to be taken into account to compute the perturbed HMI velocity. Using Eq.~\eqref{eq:lookup}, 
\begin{equation}
\vdop_{\rm HMI} = \mathcal{F}_{\rm HMI}(\vdop^0 + \delta\vdop, \mu_0 + \delta \mu) = \vdop_{\rm HMI}^0 + \delta\vdop_{\rm HMI},
\end{equation}
where the expression for $\delta\mu$ is given by Eq.~(18) in \citetalias{Kostogryz2021} and
\begin{align}
    \vdop_{\rm HMI}^0 &= \mathcal{F}_{\rm HMI}(\vdop^0, \mu_0), \label{eq:v0_HMI} \\
    \delta\vdop_{\rm HMI} &= \delta\vdop \frac{\partial \mathcal{F}_{\rm HMI}}{\partial \vdop} + \delta\mu \frac{\partial \mathcal{F}_{\rm HMI}}{\partial\mu}. \label{eq:deltav_HMI}
\end{align}
A representation of the function $\mathcal{F}_{\rm HMI}$ and its derivatives is given in Appendix~\ref{sect:lookup}. 

\subsection{Response function for velocity} \label{sect:contribution_function}

\begin{figure}[t]
    \centering
    \includegraphics[width=0.99\linewidth]{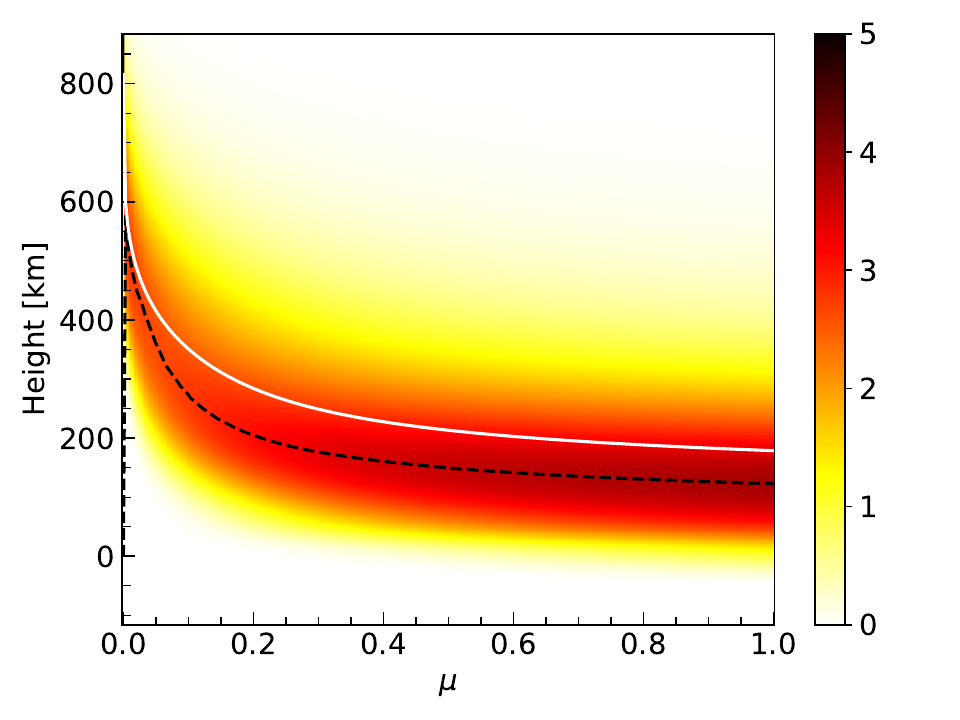}
\caption{Response function for velocity $K(s,\mu)$ (in Mm$^{-1}$) defined by Eq.~\eqref{eq:response_function_V} as a function of height $s$ and limb angle $\mu$. The black and white lines represent, respectively, the maximum and center of gravity of the response function for each value of $\mu$. }
      \label{fig:response_function_V}
\end{figure}

To simplify the computation and the interpretation of the measured velocity, a direct relationship between $\delta \vdop$ and $u_{\rm los}$ (Eq.~\eqref{eq:ulos}) would be helpful. This can be done for the term coming from $\delta I_{\rm line}$ that we will denote $\delta \vdop_{\rm line}$. Let us first rewrite $\delta I_{\rm line}$ defined in Eq.~\eqref{eq:pert_vel} as
\begin{equation}
    \delta I_{\rm line}(\lambda,\mu) = \int_0^\infty K_I(\lambda,s,\mu_0) \delta \vdop_{\rm los}(s,\mu_0) \mathrm{d}s,
\end{equation}
where $s$ is the geometrical height defined from Eq.~\eqref{eq:opticalDepth} and
\begin{equation}
    K_I(\lambda,s,\mu_0) = - \frac{\partial \log \alpha_\lambda^0}{\partial \lambda} \, \frac{\lambda}{c} \, \Bigl[ S_\lambda^0 - I_\lambda^0 \Bigr] \,  \frac{\alpha_\lambda^0}{\mu_0} \, \textrm{e}^{-\tau_\lambda^0/\mu_0}.
\end{equation}
Then, using Eq.~\eqref{eq:deltav}, we obtain
\begin{equation}
    \delta \vdop_{\rm line}(\mu_0)= \int_0^\infty K(s; \mu_0) \delta \vdop_{\rm los}(s,\mu_0) \mathrm{d}s, \label{eq:dv_line}
\end{equation}
where
\begin{equation}
   K(s; \mu_0) = \frac{\partial \mathcal{F}_{\rm HMI}}{\partial \vdop^0} \Vref \frac{\sum_{i,j} \mathcal{I}_i^0(\mu_0) \mathcal{K}_j(s; \mu)  \sin\left( 2\pi \frac{j-i}{N} \right)}{\sum_{i,j} \mathcal{I}_i^0(\mu_0)  \mathcal{I}_j^0(\mu_0) \cos\left( 2\pi \frac{j-i}{N} \right)}, \label{eq:response_function_V}
\end{equation}
with
\begin{equation}
    \mathcal{K}_j(s; \mu_0)  = \int_{-\infty}^{\infty} F_j(\lambda) \left[ K_I(s,\mu_0) \ast G_{\rm macro}(\mu_0) \right]_\lambda \mathrm{d}\lambda. \label{eq:response_function_Ij}
\end{equation}
A representation of this kernel $K$ is given in Fig.~\ref{fig:response_function_V} shows that the observed velocity corresponds to a weighted average of the oscillation velocity at different depths. This averaging depends on the center-to-limb distance, particularly for small values of $\mu$ ($\mu \leq 0.2$). Close to the disk center, the maximum is obtained around 120~km above the surface and the center-of-gravity is at 180~km. These values are in agreement with the formation heights reported by \citet{Fleck2011} and \citet{Nagashima2014} using the HMI algorithm on 3D radiation-hydrodynamic simulations. However, this function is far from a delta function (with height), and the observed velocity corresponds to a weighted averaging from the surface to about 400~km above the surface at disk center. A similar approach was undertaken by \citet{Vukadinovic2022_rf} to construct a kernel that relates the weak-field estimate of the line-of-sight magnetic field to the underlying depth-dependent field. Finally, we would like to note that Eq.~\eqref{eq:response_function_Ij} can be used to define a response function and a formation height for the six measured HMI intensities.

\subsection{Perturbed continuum intensity}

The perturbed continuum intensity is obtained in a similar way to velocity. In Appendix~\ref{sect:Ic}, we derived 
\begin{align}
    \delta I_{\rm HMI} = \frac{1}{N} \sum_{j=1}^N \delta\mathcal{I}_j +  \mathrm{e}^{- \frac{( \lambda_j - \lambda_{\rm HMI} )^2}{\sigma_0^2}} \left(\delta I_d   + 2 I_d^0 \frac{\delta\sigma}{\sigma_0} \frac{(\lambda_j - \lambda_{\rm HMI})^2}{\sigma_0^2}  \right)  , \label{eq:deltaIc}
\end{align}
where the perturbed line width and line depth are
\begin{align}
    \delta\sigma &= \frac{T^2}{6\pi^2\sigma_0} \left[ \frac{a_1^0 \delta a_1 +  b_1^0 \delta b_1}{(a_1^0)^2 + (b_1^0)^2 } - \frac{a_2^0 \delta a_2 +  b_2^0 \delta b_2}{(a_2^0)^2 + (b_2^0)^2 } \right], \\
    \delta I_d &= I_d^0 \left[- \frac{\delta\sigma}{\sigma_0} +  \frac{\delta a_1 a_1^0+\delta b_1 b_1^0}{(a_1^0)^2+(b_1^0)^2} + \frac{2\pi^2\sigma_0 \delta\sigma}{T^2} \right]. 
\end{align}
Note that the line width and line depth are also HMI observables (hmi.Lw\_45s and hmi.Ld\_45s respectively) but we did not study them in detail here.

\subsection{Validation}
To assess whether the intensity perturbation using first-order perturbation theory is functioning correctly, we first present a validation of our algorithm. We perturb the background model from Sect.~\ref{sect:background} characterized by $(r_0,p_0,T_0)$ (and intensity $I_\lambda^0$), using a simple function representative of a radial eigenfunction. We then compute the intensity $I^\lambda(\mu)$ at each wavelength in the perturbed medium, which has a surface given by $r_0 \mathbf{e}_{\rm r} + \bxi$, velocity $\mathrm{d}_t \bxi$, pressure $p = p_0 + \delta p$ and temperature $T = T_0 + \delta T$. The difference between the perturbed and background intensity $I_\lambda - I_\lambda^0$ is then compared to the first-order computation of $\delta I_\lambda$ derived in Sect.~\ref{sect:firstOrderI}. 

\begin{figure}[t]
    \centering
    \includegraphics[width=0.99\linewidth]{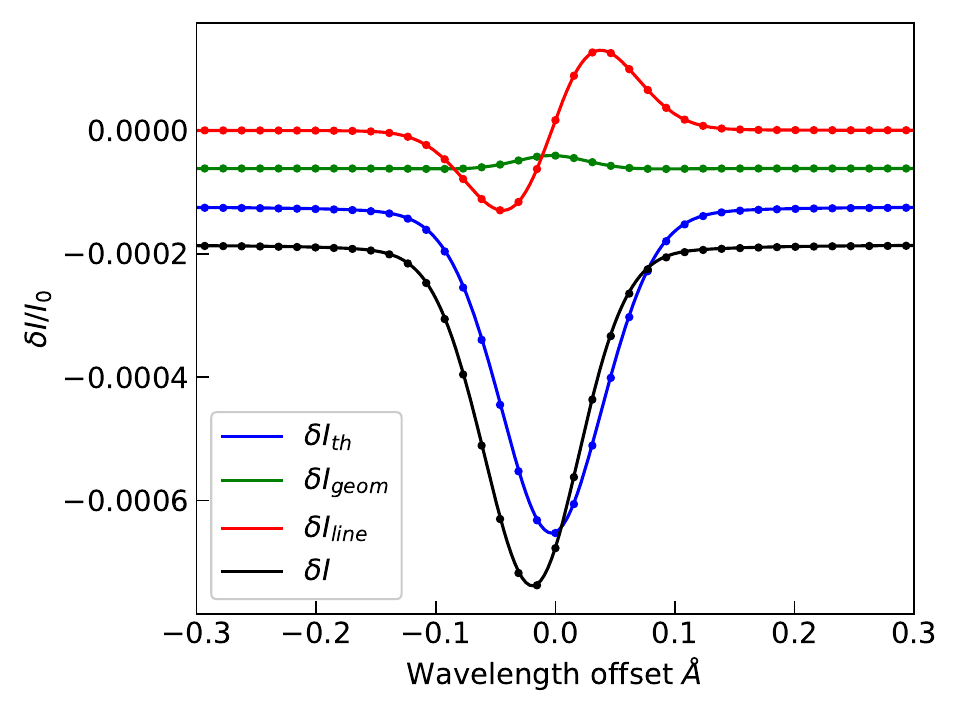}
\caption{Test of the direct computation of the perturbed intensity $(I_\lambda-I_\lambda^0)/I_\lambda^0$ (dots) for $\mu=0.5$ and comparison with first-order computations $\delta I_\lambda / I_\lambda^0$ (solid line). The thermodynamic, geometrical, and line contributions are represented separately in blue, green, and red, respectively, while the full intensity is in black. The wavelength offset is with respect to the center of the HMI line $\lambda_{\rm HMI} = 6173.33$~\AA. }
      \label{fig:test_deltaI_HMI}
\end{figure}

In Fig.~\ref{fig:test_deltaI_HMI}, we compare the different contributions separately, for example, $I_{\rm th}$ is computed in a background medium with perturbed pressure and temperature but without surface deformation or velocity. The direct and first-order computations agree very well for the different contributions, which allows us to validate the derivation of  Sect.~\ref{sect:firstOrderI} and the numerical implementation. 

Once the intensity is known at all wavelengths, the continuum intensity and the velocity can be computed following the HMI algorithm. A comparison of these two observables, obtained from both the direct computation and the first-order formulation, is presented in Appendix~\ref{sect:validation}, Figs.~\ref{fig:test_deltaV} and \ref{fig:test_continuum_HMI} for the HMI velocity and continuum intensity. Both methods agree well validating the expressions of $\delta \vdop$ (Eq.~\eqref{eq:deltav}) and $\delta I_{\rm HMI}$ (Eq.~\eqref{eq:deltaIc}).

\section{HMI line-of-sight Doppler velocity} \label{sect:velocity}

The interpretation of Doppler velocity often relies on the assumption that the observed Doppler velocity is simply a line-of-sight projection at a given height. In this section, we study how our more comprehensive approach differs from this simple approximation. We consider the different terms that enter into the computation of the HMI velocity
\begin{equation}
    \delta \vdop_{\rm HMI} = \delta \vdop_{\rm line} + \delta \vdop_{\rm th} + \delta \vdop_{\rm geom}. \label{eq:delta_vhmi}
\end{equation}

The framework developed here is general and could be used with any perturbation. Here, we perturb our atmosphere by a single mode $\bxi_{nl}$ associated with the eigenfrequency $\omega_{nl}$ and compute the resulting velocity $\delta \vdop_{\rm HMI}$. The eigenfunctions are computed in a (non-rotating) spherically symmetric background and thus do not depend on the longitudinal wavenumber $m$. We use the patched model between \texttt{Model~S} and the \texttt{MPS-ATLAS} atmosphere presented in Sect.~\ref{sect:background_model} and compute the adiabatic eigenfunctions using the GYRE code \citep{Townsend2013}.  We normalize the eigenfunction so that the line-of-sight velocity has an amplitude of 1.0~m/s at the disk center at the surface. For an eigenfunction $\bxi_{nl}(\br,t) = \Re[\bxi_{nl}(\br) {\rm e}^{\ii \omega_{nl} t}]$, we compute the (complex) velocity $\delta \vdop_{\rm HMI}(\br)$ and the associated contributions $\delta \vdop_{\rm line}$, $\delta \vdop_{\rm th}$, and $\delta \vdop_{\rm geom}$. The final time-dependent quantities are then obtained by multiplying by ${\rm e}^{\ii\omega_{nl}t}$ and taking the real part. 

Under our hypotheses, $\bxi_{nl}$ is real, which implies that the line contribution $\delta \vdop_{\rm line}$ is purely imaginary while the geometrical and thermodynamical terms are real.  We will thus compare $\delta \vdop_{\rm line}$ to
\begin{equation}
   \delta\vdop_{\rm los}(h) = -\ii \omega \ \mathbf{\mathrm{e}}_{\rm obs} \cdot \bxi(h) \label{eq:delav_los}
\end{equation}
 and $\delta \vdop_{\rm geom}, \delta \vdop_{\rm th}$ to 0. Two line-of-sight approximations are considered: once at the center-of-gravity of the response function at each $\mu$ position $h_\mu$ and once at a fixed height, independent of the position on the disk, corresponding to $h_{\mu=1} = 180$~km (see Fig.~\ref{fig:response_function_V}).

\subsection{Velocity perturbation caused by a radial mode} \label{sect:vel_radial}

\begin{figure}[t]
    \centering
    \includegraphics[width=0.99\linewidth]{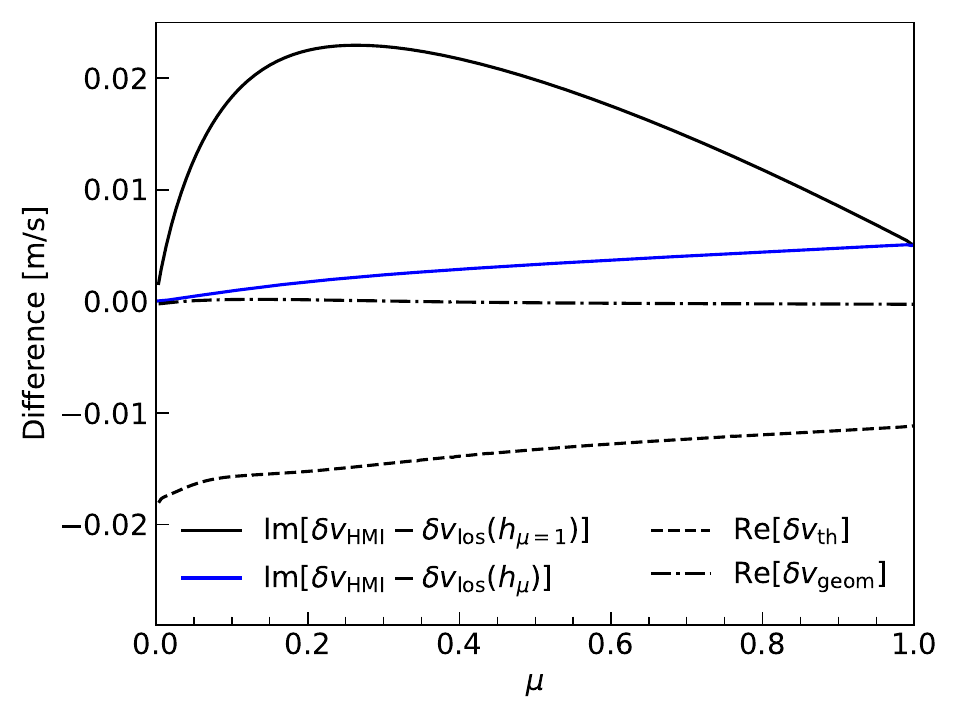}
\caption{Difference between the velocity perturbation caused by an example radial p-mode ($l =0, n=20, \omega_{nl}/2\pi = 2.90$~mHz) of amplitude 1.0~m/s (at the surface at the disk center) computed from the HMI algorithm (without background velocity) and a simple line-of-sight approximation evaluated at the formation height at each position on the disk $h_\mu$ or at disk center $h_{\mu=1}$. Note that $\delta \vdop_{\rm los}$ is purely imaginary so that Re[$\delta \vdop_{\rm th}$] and Re[$\delta \vdop_{\rm geom}$] are a deviation from the simple line-of-sight approximation. A center-to-limb effect is observed in the imaginary part and in the thermodynamic contribution. For this radial mode, the geometrical component is small.}
      \label{fig:velocity_pmode_all}
\end{figure}

Figure~\ref{fig:velocity_pmode_all} studies the different contributions to $\delta \vdop_{\rm HMI}$ when the perturbation is caused by the oscillation of a single radial p-mode with ($l =0, n=20, \omega_{nl}/2\pi = 2.90$~mHz).  As expected, the difference between $\delta \vdop_{\rm line}$ and the single height line-of-sight approximation is increasing toward the limb as the line forms higher in the atmosphere when moving toward the limb. For this contribution, the error can be decreased by using the formation height $h_\mu$ that depends on the center-to-limb distance. By doing so, the absolute error is about 0.6\% and remains mostly constant with $\mu$. We note that, by definition of the center of gravity, this error is zero if the eigenfunction varies linearly with height.

Another contribution due to the structural changes (mostly temperature) causes a non-zero real part. This term modifies the shape of the line symmetrically compared to the central wavelength and should thus not contribute to the velocity. This is, unfortunately, not the case as the filters and the line shape (not considered in this study) are not symmetric.
This thermodynamical contribution causes a systematic phase shift with nontrivial center-to-limb variations as it is out of phase compared to $\delta \vdop_{\rm line}$.

\subsection{Velocity perturbation caused by high-degree modes}

\begin{figure*}
\begin{tabular}{cc}
\includegraphics[width=0.47\linewidth]{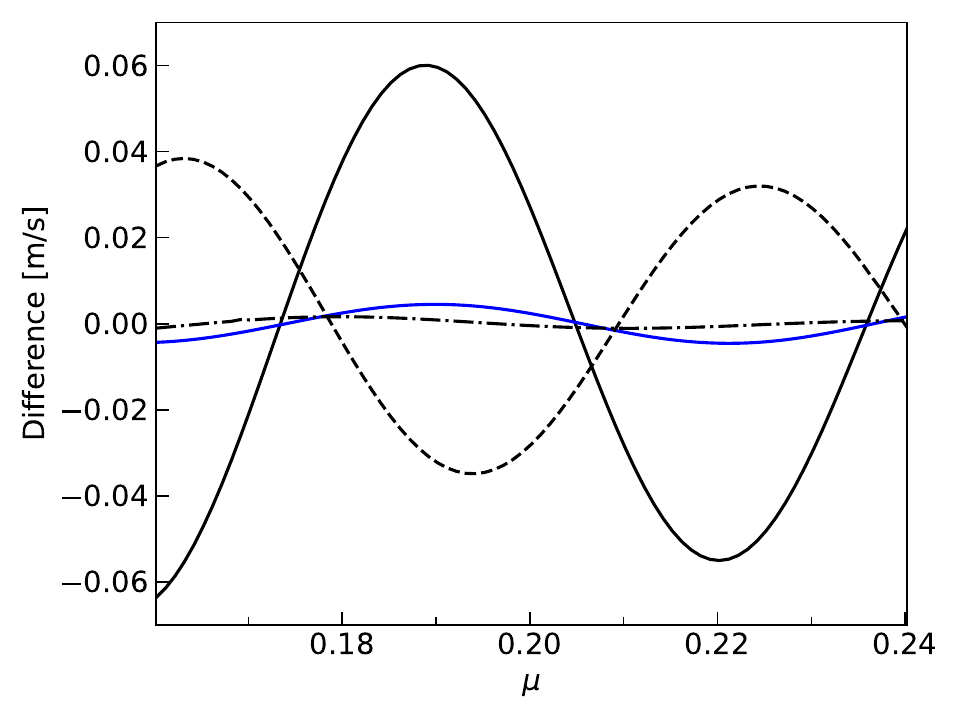} & \includegraphics[width=0.47\linewidth]{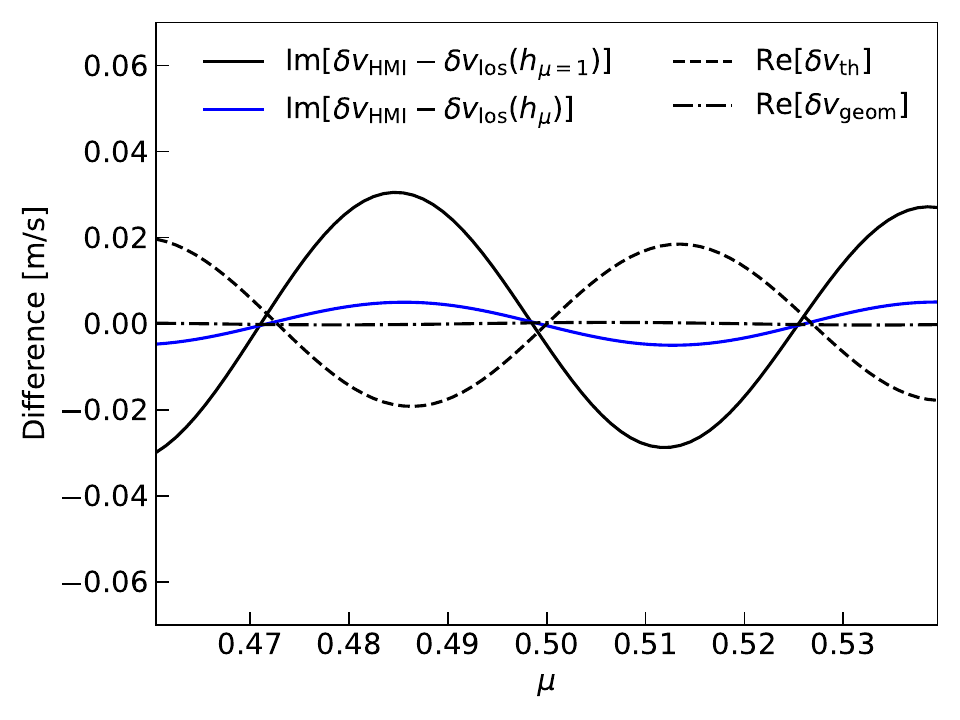} 
\\
\includegraphics[width=0.47\linewidth]{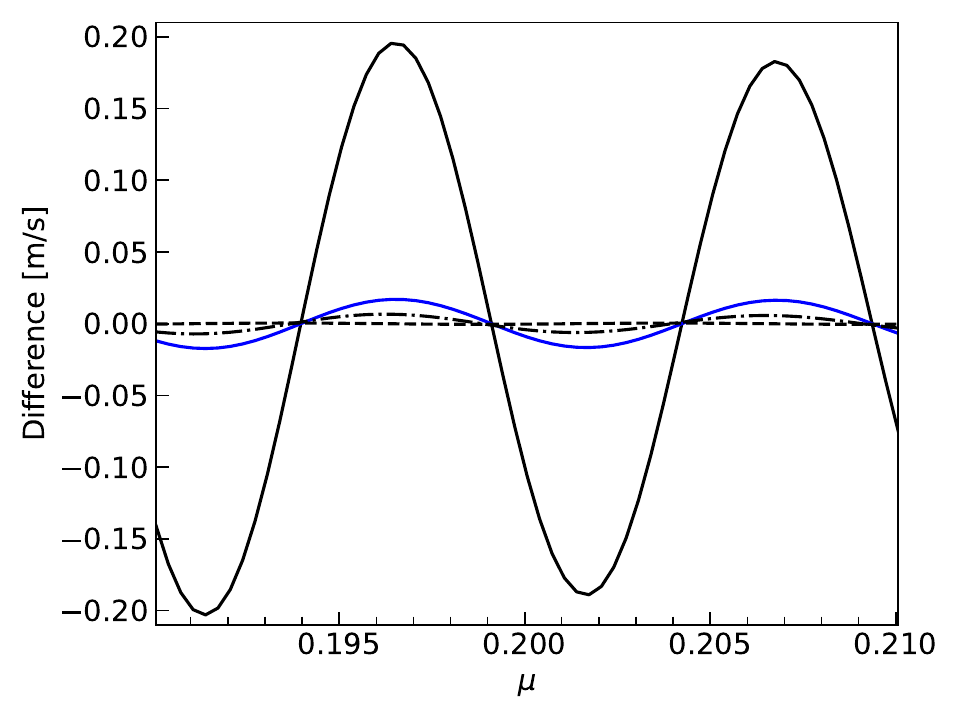} & \includegraphics[width=0.47\linewidth]{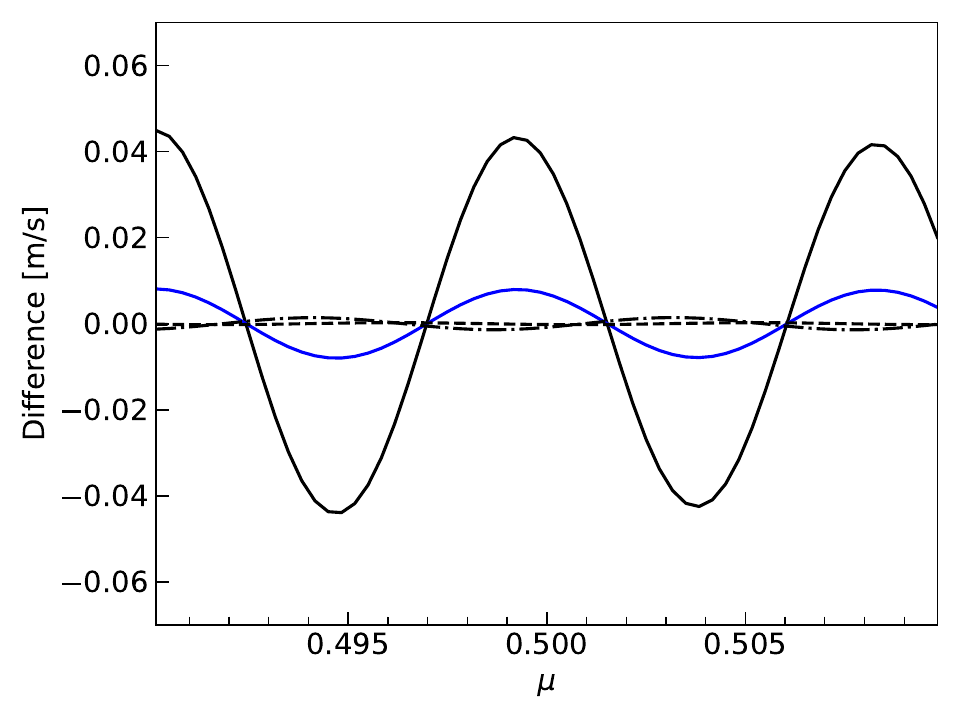}
\end{tabular}
\caption{Same as Fig.~\ref{fig:velocity_pmode_all} for an example high-degree p-mode ($l = 100, n=6, \omega_{nl} / 2\pi = 2.93$~mHz) (top) and the f-mode ($l = 600, n=0, \omega_{nl} / 2\pi = 2.45$~mHz) (bottom) centered around $\mu=0.2$ (left) and $\mu=0.5$ (right) on the central meridian. 
}
\label{fig:velocity_pmode_high}
\end{figure*}

Figure~\ref{fig:velocity_pmode_high} shows the perturbed velocity caused by a p-mode with $l=100$ and $n=6$ corresponding to a frequency of 2.93~mHz and a f-mode with $l=600$ (frequency 2.45~mHz). As observed for the radial mode, the error due to the formation height is increasing toward the limb. This error is also increasing with the harmonic degree. The geometrical contributions are also increasing with the harmonic degree but remain relatively small. As expected, the thermodynamic effect vanishes for the f-mode.  Overall, for all modes, thermodynamic and/or geometrical contributions that are out of phase with the Doppler contributions will cause systematic center-to-limb phase shifts. 
We note that these contributions vanish if we artificially symmetrize the HMI filters around the central wavelength (for a given background velocity). They are thus due to the shift of the line due to the background velocity resulting in asymmetric filters with respect to the HMI central wavelength. The (instrumental) asymmetry of the filters (and eventually of the line shape) also contributes to this phase shift.

\subsection{Modeling on the CCD with realistic background velocities}

\begin{figure*}[!htb]
    \centering
    \includegraphics[width=0.99\linewidth]{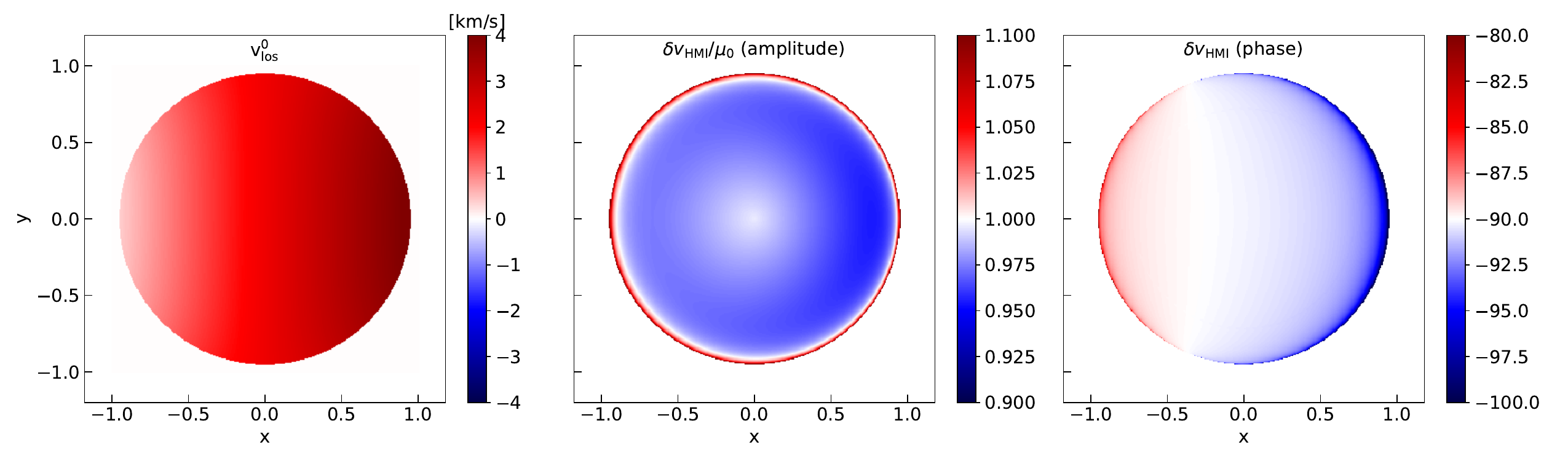} 
 \caption{Effect of background velocity on the perturbed velocity caused by an example radial mode ($l =0, n=20, \omega_{nl}/2\pi = 2.90$~mHz) on June, 6th, 2011 at 00:00 when the $B_0$ angle is close to 0. Left:  Line-of-sight component of the background velocity due to the satellite motion and solar differential rotation. 
 The middle panel shows the amplitude (normalized by $\mu_0$) and the right panel corresponds to the phase (in degrees) of the perturbed velocity computed using the HMI algorithm. A one-day animation of the observations is available at \href{https://edmond.mpg.de/dataset.xhtml?persistentId=doi:10.17617/3.FBBGMH}{https://edmond.mpg.de/dataset.xhtml?persistentId=doi:10.17617/3.FBBGMH}, illustrating the variations caused by the changing spacecraft motion. }
      \label{fig:deltav_obsvr}
\end{figure*}

To get closer to the observations, we model Doppler velocity on the CCD as it would be observed by HMI and include the effects of background velocities, in particular the satellite velocity and differential rotation. We want to study the impact of background velocities on the evaluation of the perturbed velocities.

We denote by $(x,y)$ the pixel coordinates on the CCD. We take into account the B$_0$- and P$_0$-angles in order to make the connection between the CCD coordinates $(x,y)$ and the heliographic angles $(\theta,\phi)$. To do so, we use the relations between pseudo-angles and the TAN projection as described in Sect. 7.2 of \citet{Thompson2006}. The different values required to make the conversions are directly read from the HMI header.

We also include the satellite velocity associated to a given time frame by reading the keywords OBS\_VW, OBS\_VN, and OBS\_VR from the HMI header. The background velocity is the sum of line-of-sight projection of the SDO motion \citep[see for example Eq. 4 in][]{Schuck2016} and the differential rotation given by
\begin{equation}
    \vdop_{\rm los}^{\rm rot}(\theta,\phi) = \mathbf{u}^{\rm rot} \cdot \mathrm{e}_{\rm obs} = - R_\odot\Omega(\theta)  \sin\theta \cos B_0 \sin\phi.
\end{equation}
We use a three-term approximation of the surface solar differential rotation  $\Omega(\theta) / 2\pi = [454 - 55 \cos^2\theta - 76 \cos^4\theta]$~nHz.

\begin{figure*}[t]
    \centering
\includegraphics[width=0.99\linewidth]{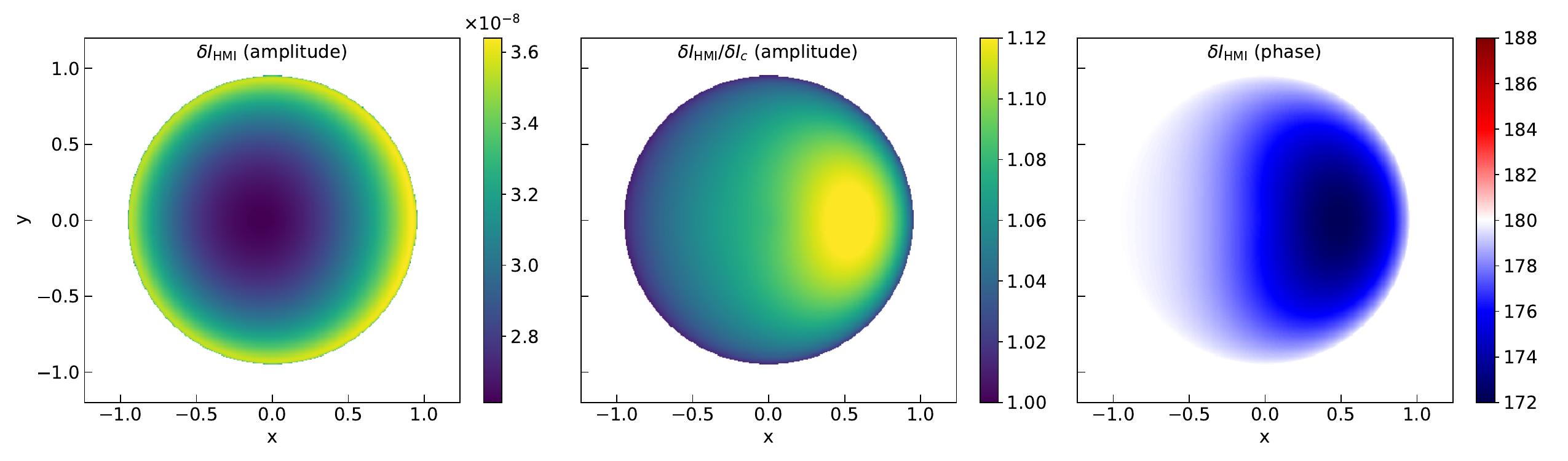} 
\caption{Continuum intensity perturbation caused by a radial p-mode ($l =0, n=20, \omega_{nl}/2\pi=2.90$~mHz) on June, 6th, 2011 at 00:00. Amplitude (left) and phase (right) of the perturbed intensity computed using the HMI algorithm. The middle panel shows the ratio between HMI and the theoretical continuum. An animation over one day of observation is available at \href{https://edmond.mpg.de/dataset.xhtml?persistentId=doi:10.17617/3.FBBGMH}{https://edmond.mpg.de/dataset.xhtml?persistentId=doi:10.17617/3.FBBGMH}.}
    \label{fig:continuum}
\end{figure*}

A representation of the background velocity for a particular frame on June, $6^\textrm{th}$ 2011 at 00:00 is shown on the left panel of Fig.~\ref{fig:deltav_obsvr}. This day was chosen as the $B_0$-angle is small. On this frame, the background velocity varies between 0 and 4~km/s, with the smallest values on the East where the satellite motion almost compensates the rotation. Additional background flows such as convective blue shift and gravitational red shift could be added within this framework.

The middle and right panel of Fig.~\ref{fig:deltav_obsvr} shows the normalized amplitude ($\delta\vdop_{\rm HMI} / \mu_0$) and the phase of the perturbed velocity caused by a radial mode. Such phase maps are often used to study systematics \citep[see for example][]{Couvidat2016}. They are computed on a CCD grid with 300 pixels uniformly distributed in the x- and y- directions. From a simple line-of-sight velocity projection, the amplitude should be constant and equal to one over the disk and the phase should be $-90^\circ$. The variations in amplitude over the disk are mostly due to the Doppler shift term $\delta \vdop_{\rm line}$ and depends on the center-to-limb distance. However, these variations also depend on the background velocity (see also the online movie associated with Fig.~\ref{fig:deltav_obsvr} showing the 24-hour variations of these quantities).  The phase is mostly created by the thermodynamical contributions (the geometrical part is about one order of magnitude smaller, but becomes more significant for higher-degree modes) and is anticorrelated to the background rotation with a Pearson correlation coefficient of -0.8.  We note that the imaginary part of $\delta\vdop_{\rm HMI}$ is almost completely anticorrelated to the rotation (correlation coefficient of -0.98).

\section{HMI Continuum Intensity} \label{sect:intensity}

The continuum intensity derived from the HMI algorithm (\Ihmi) is underestimated in the quiet Sun \citep{Couvidat2012}. This underestimation arises due to the imperfections of the HMI algorithm which assumes a Gaussian line profile and employs non-ideal $\delta$-function filters. These limitations leads to inaccurate discrete approximations of Fourier coefficients.
Furthermore, the Doppler effect caused by the SDO's motion around the Earth and the Sun results in a shift of the spectral line with respect to the HMI filters, contributing differently to each filtergram.

Figure~\ref{fig:continuum} shows the perturbed continuum intensity as computed by the HMI algorithm  for a radial p-mode ($l =0, n=20, \omega_{nl}/2\pi = 2.90$~mHz). As was already noted for the theoretical continuum, the intensity depends on the distance to the disk center \citep{Toutain1999,Kostogryz2021}. However, the HMI continuum differs from the theoretical one by more than 10\% depending on the position on the disk. Both the phase and the amplitude are affected with non-trivial variations across the detector. The thermodynamic contributions are mostly responsible for the amplitude variations while the phase is linked to the imaginary contribution due to the line shift.

\section{Summary and discussion} \label{sect:discussion}

In this paper, we outlined the steps required to compute the perturbations in continuum intensity and Doppler velocity caused by oscillations of the solar surface. We found that approximating \Ihmi \ by the theoretical continuum and \Vhmi \ by a line-of-sight projection of the oscillations lead to an amplitude error around 10\% and a phase error up to $10^\circ$.  Due to the asymmetry of the filters with respect to the central wavelength (caused by background velocities as well as the filters themselves), \Vhmi\  is also influenced by thermodynamic perturbations ($\delta\vdop_{\rm th}$), while oscillation velocities ($\delta\vdop_{\rm line}$) also contribute to \Ihmi. A polynomial correction is implemented in the HMI pipeline to take into account the background velocities \citep{Couvidat2016} and the quality of this correction could be assessed using the tools developed in this paper. Other sources of background velocities, such as convective blue shift and gravitational red shift \citep[see for example][]{Beckers1978} should be added to the background velocity before comparing to observations.  

More generally, the framework developed here can be used to gain insight into the systematic errors reported in the different helioseismic analyses. Understanding these errors can also allow us to make use of additional measurements, such as the frequency-filtered travel times from \citet{Rajaguru2020} or the multi-height measurements proposed by \citet{Nagashima2014}. Note however that this framework cannot explain (instrumental) long-term variations as observed in the travel times by \citet{Liang2018} (their Figure~4). The expressions for the perturbed intensity and velocity can also be used to construct improved leakage matrices \citep[see for example][]{Larson2015} for global helioseismology or normal-mode coupling.

The framework has been illustrated on HMI but can be applied to other instruments such as PHI and PMI. By shifting the spectral line to the Ni~I~6768 line and adapting the algorithms, the data from MDI and GONG can also be analyzed, allowing for a comparison of systematic errors between the different instruments.

\begin{acknowledgements}
We thank Philip Scherrer and Todd Hoeksema for useful discussions about the HMI algorithm and HMI observables and the provision of example HMI filters. We thank  Zhi-Chao Liang for help with HMI keywords and the projection algorithm, as well as Aaron Birch and Ha~Pham for comments on an earlier version of this manuscript.
NK and LG acknowledge generous support from the  German Aerospace Center (DLR) under grants ``PLATO Data Center'' 50OO1501 and 50OP1902. 
DF and LG acknowledge partial support from ERC Synergy grant WHOLE~SUN 810218, from DFG grant  SFB 1456 ``Mathematics of Experiments'' (project C04), and from DFG grant ``Stellare Schmetterlingsdiagramme'' no. 530101854. 
VW and AIS acknowledge support from ERC Synergy Grant REVEAL under the European Union’s Horizon 2020 research and innovation program (grant no. 101118581) IM acknowledges the financial support from the Serbian Ministry of Science and Technology through the grants 451-03-136/2025-03/200104 and 451-03-136/2025-03/200002.
\end{acknowledgements}

\bibliographystyle{aa}
\bibliography{References}

\clearpage

\appendix

\section{Validation} \label{sect:validation}

This appendix shows the validation of the computation of perturbed continuum intensity and velocity. To do so, we compute the intensity at each wavelength in a background model characterized by $(r_0,p_0,T_0,\bu=0)$ and perturbed intensities. The perturbed intensities are computed with a media characterized by $(r_0,p,T,\bu=0)$ for $\delta I_{\rm th}$, $(r_0 \mathbf{e}_{\rm r} + \bxi, p_0, T_0,\bu=0)$ for $\delta I_{\rm geom}$, and  $(r_0,p_0,T_0, \mathrm{d}_t \bxi)$ for $\delta I_{\rm line}$.

\begin{figure}[h]
    \centering
    \includegraphics[width=0.83\linewidth]{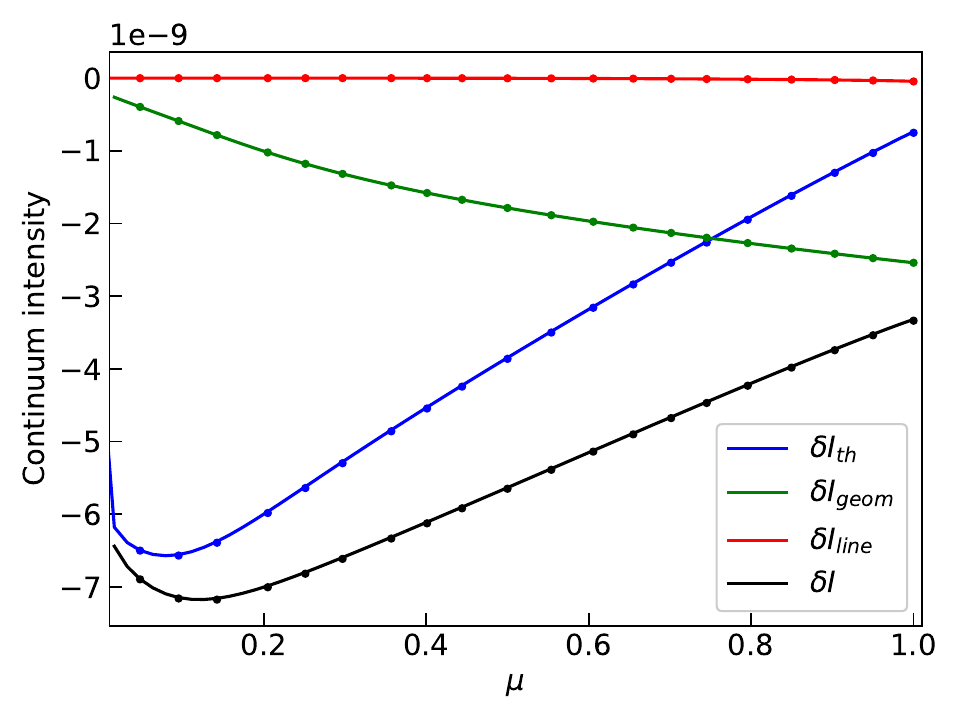}
\caption{Test of the direct computation of continuum intensity (dots) as a function $\mu$ and comparison with the first-order computation (solid line). The thermodynamical, geometrical, and line contributions are represented separately in blue, green, and red respectively while the full intensity is in black. The wavelength offset is with respect to the center of the HMI line $\lambda_{\rm HMI} = 6173.33$~\AA. }
      \label{fig:test_continuum_HMI}
\end{figure}

\begin{figure}[]
    \centering
    \includegraphics[width=0.83\linewidth, clip, trim=0 0 0 0.3cm]{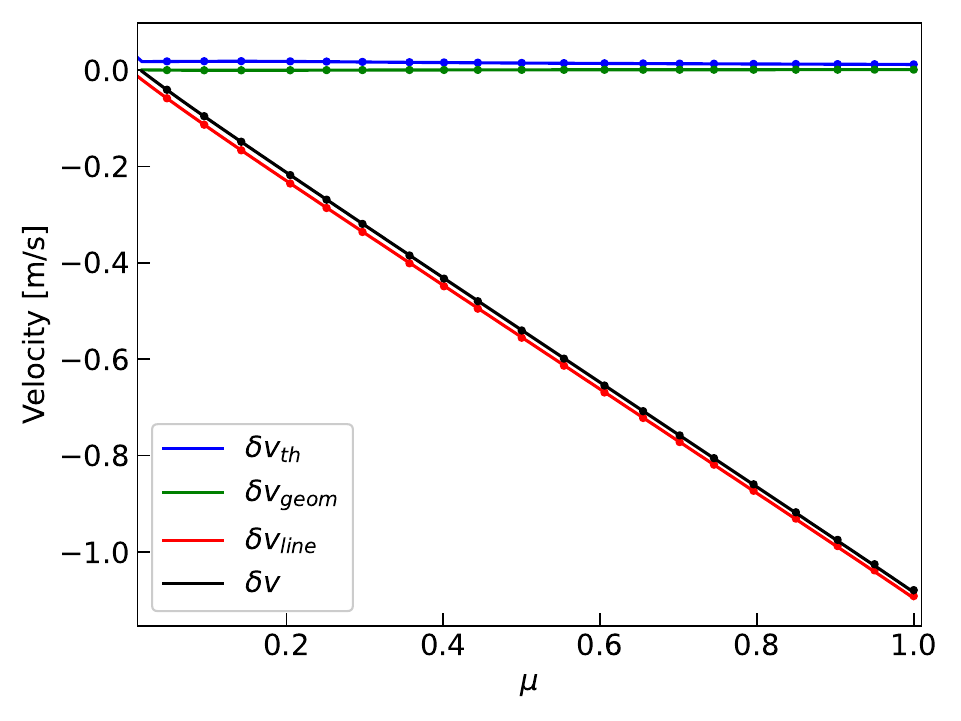}
\caption{Test of the direct computation of velocity and comparison with the first-order computation. The velocity computation is done with the MDI-like algorithm after convolution of the intensity with the filtergrams shown in Fig.~\ref{fig:I0}. The different contributions as well as the total velocity are shown using the first-order approach (solid line) and direct computation (dots).}
      \label{fig:test_deltaV}
\end{figure}

A validation of the intensity as a function of wavelength in shown in the main text in Fig.~\ref{fig:test_deltaI_HMI}. Here, we additionally compare the perturbed continuum intensity in Fig.~\ref{fig:test_continuum_HMI} and velocity in Fig.~\ref{fig:test_deltaV}. The direct and first-order computations agree well. For velocity computations, the main contribution is coming as expected from the line shift (that is the term $\delta I_{\rm line}$) but small deviation due to the thermodynamic term is observed in particular for small values of $\mu$. Such a deviation could introduce systematic effects in the interpretation of the observables.


\FloatBarrier

\section{Look-up tables} \label{sect:lookup}

\begin{figure*}[btp]
    \centering
    \includegraphics[width=0.99\linewidth]{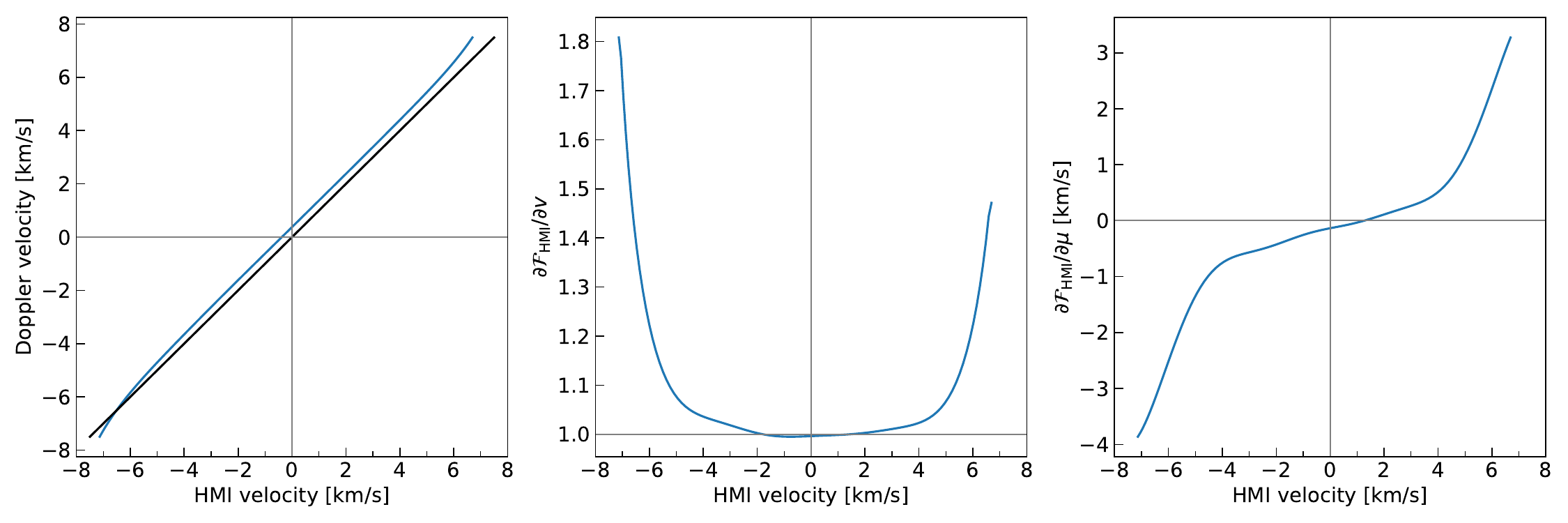}
\caption{Look-up table and derivatives at $\mu=1$ necessary to map the HMI velocity to real Doppler velocity. Left: $\mathcal{F}_{\rm HMI}$ (blue line) compared to the one-to-one correspondence (black). Middle:  $\partial \mathcal{F}_{\rm HMI} / \partial \vdop^0$. Right: $\partial \mathcal{F}_{\rm HMI} / \partial \mu_0$. These two derivatives are necessary to compute the perturbed velocity.}
\label{fig:lookup_equator}
\end{figure*}

The look-up table gives the correspondence between the velocity computed from the HMI algorithm and the real Doppler velocity 
\begin{equation}
    \vdop^0_{\rm HMI} = \mathcal{F}_{\rm HMI}(\vdop^0, \mu_0).
\end{equation}
To obtain $\mathcal{F}_{\rm HMI}$, we manually shift the line of a given wavelength $\Delta\lambda$ corresponding to a Doppler shift of $\Delta\lambda \, c / \lambda$ and compare this shift to $\vdop^0_{\rm HMI}$ using the HMI algorithm. A representation of this function at the disk center is given on the left panel of Fig.~\ref{fig:lookup_equator}. As the filters are not symmetric, an offset is visible between the velocity obtained by the algorithm and the real velocity. Otherwise, the slope is very close to 1 with some deviations only for very large background velocities (larger than 6~km/s).

To compute the perturbed velocity, we also need the derivatives of the look-up table with respect to $\vdop^0$ and $\mu_0$ as 
\begin{equation}
    \delta\vdop_{\rm HMI} = \delta\vdop \frac{\partial \mathcal{F}_{\rm HMI}}{\partial \vdop^0} + \delta\mu \frac{\partial \mathcal{F}_{\rm HMI}}{\partial\mu_0}. \label{eq:look_up_dv}
\end{equation}
A representation of these two derivatives is shown in Fig.~\ref{fig:lookup_equator}. For small background velocities $\frac{\partial \mathcal{F}_{\rm HMI}}{\partial \vdop^0} \approx 1$ and $\frac{\partial \mathcal{F}_{\rm HMI}}{\partial\mu_0} \approx 0$ so that $\delta\vdop_{\rm HMI} \approx \delta\vdop$. However, some strong deviations appear for large background velocities which could lead to an underestimation of the real velocity (as $\frac{\partial \mathcal{F}_{\rm HMI}}{\partial \vdop^0} > 1$) but also to some variations depending on the distance to the disk center when $\frac{\partial \mathcal{F}_{\rm HMI}}{\partial\mu_0}$ is significant.

\begin{figure}[!htb]
    \centering
    \includegraphics[width=0.99\linewidth]{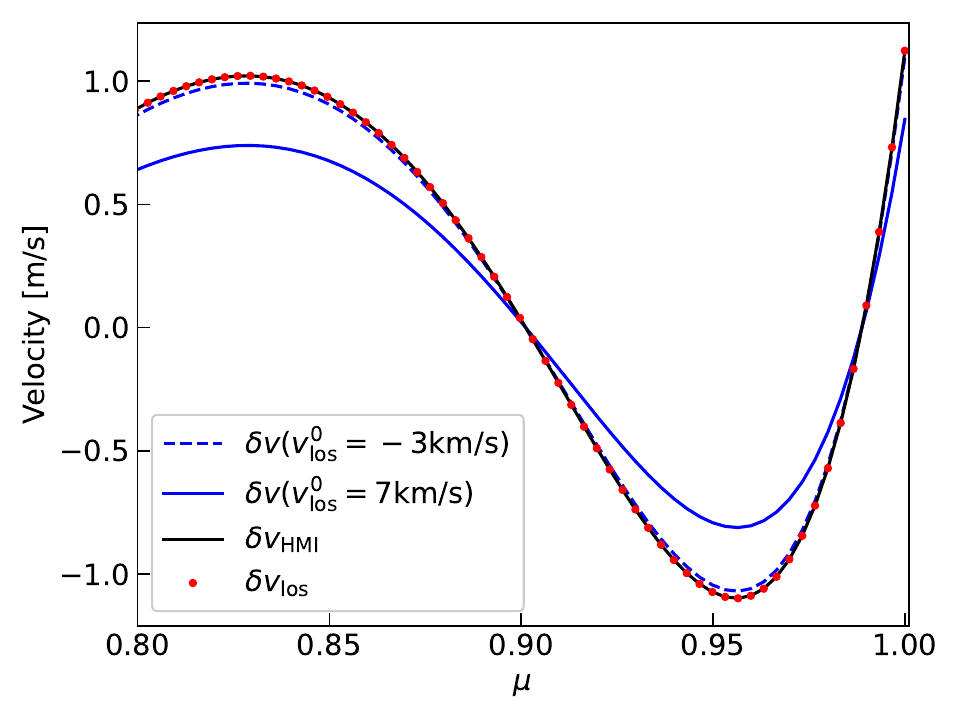}
\caption{Perturbed velocity computed with the HMI algorithm with and without applying the look-up table. For large background velocities, the amplitude correction due to the look-up table is significant. Note that $\delta \vdop_{\rm HMI}$ is the same for all background velocities.}
\label{fig:test_lookup}
\end{figure}

To show the importance of Eq.~\eqref{eq:look_up_dv} in computing the perturbed velocity, we perturb the reference model by a function that is independent of depth (and varies with latitude as a Legendre polynomial of order 10). In this case, we know the exact value of the perturbed velocity and we can compare it to the value returned by the HMI algorithm with or without the look-up tables. Such a comparison is made in Fig.~\ref{fig:test_lookup}. After taking the look-up table into account, the velocity $\delta \vdop_{\rm HMI}$ perfectly matches the expected value. However, it deviates from the value without applying the look-up table, in particular for very large background velocities. We note finally that the correction due to the term $\delta\mu \frac{\partial \mathcal{F}_{\rm HMI}}{\partial\mu_0}$ is very small (5 order of magnitude smaller than the other term in this case).

\onecolumn

\section{First-order perturbations in velocity and intensity} \label{sect:first_order}

In order to obtain the first-order perturbations in the different HMI observables, we decompose the Fourier coefficients $a_1$ and $b_1$ as
\begin{equation}
    a_1 = a_1^0 + \delta a_1 \quad \textrm{and} \quad b_1 = b_1^0 + \delta b_1,
\end{equation}
where $a_1^0$ and $b_1^0$ (resp. $\delta a_1$ and $\delta b_1$) are computed from Eqs.~\eqref{eq:a1} and \eqref{eq:b1} using the background intensity $\mathcal{I}_j^0$ (resp. perturbed intensities $\delta \mathcal{I}_j$).

\subsection{Velocity} \label{sect:first_order_vel}

From the velocity definition (Eq.~\eqref{eq:velocity}), we obtain
\begin{equation}
    \vdop = \Vref \arctan \left( \frac{b_1^0 + \delta b_1}{a_1^0 + \delta a_1} \right)
    = \Vref \arctan \left( \frac{b_1^0}{a_1^0} + \frac{\delta b_1}{a_1^0} - \frac{\delta a_1 b_1^0}{(a_1^0)^2} \right) .
\end{equation}
We use a first order development of the arctan
\begin{equation}
    \arctan(x+\epsilon) = \arctan x + \frac{\epsilon}{1+x^2},
\end{equation}
to obtain $\vdop = \vdop^0 + \delta \vdop$ where
\begin{align}
\vdop^0 &= \Vref \arctan \left( \frac{b_1^0}{a_1^0} \right), \\
    \delta \vdop &= \Vref \frac{a_1^0 \delta b_1 - b_1^0 \delta a_1}{(a_1^0)^2 + (b_1^0)^2}.
\end{align}
Replacing by the expressions of the Fourier coefficients, $\delta \vdop$ can also be written as 
\begin{equation}
    \delta \vdop = \Vref \frac{\sum_{i,j} \mathcal{I}_i^0 \delta \mathcal{I}_j \sin\left( 2\pi \frac{j-i}{N} \right)}{\sum_{i,j} \mathcal{I}_i^0  \mathcal{I}_j^0 \cos\left( 2\pi \frac{j-i}{N} \right)}. 
\end{equation}

\subsection{Line width}

Using the first-order perturbation of the Fourier coefficients, the line width $\sigma$ is rewritten as
\begin{align}
\sigma &:= \frac{T}{\pi \sqrt{6}} \sqrt{\log \left( \frac{a_1^2 + b_1^2}{a_2^2 + b_2^2} \right)}
     = \frac{T}{\pi \sqrt{6}} \sqrt{\log \left[(a_1^0)^2 + (b_1^0)^2 + 2a_1^0 \delta a_1 + 2 b_1^0 \delta b_1 \right] - \log \left[(a_2^0)^2 + (b_2^0)^2 + 2a_2^0 \delta a_2 + 2 b_2^0 \delta b_2 \right] }, \\
    &= \frac{T}{\pi \sqrt{6}} \sqrt{\log \left[(a_1^0)^2 + (b_1^0)^2 \right] + 2 \frac{a_1^0 \delta a_1 +  b_1^0 \delta b_1}{(a_1^0)^2 + (b_1^0)^2 }  - \log \left[(a_2^0)^2 + (b_2^0)^2 \right] - 2 \frac{a_2^0 \delta a_2 +  b_2^0 \delta b_2}{(a_2^0)^2 + (b_2^0)^2 }}.
\end{align}
We obtain $\sigma = \sigma_0 + \delta \sigma$ where the perturbation to the line width $\delta \sigma$ is given by
\begin{equation}
  \delta\sigma =  \frac{T^2}{6\pi^2\sigma_0} \left[ \frac{a_1^0 \delta a_1 +  b_1^0 \delta b_1}{(a_1^0)^2 + (b_1^0)^2 } - \frac{a_2^0 \delta a_2 +  b_2^0 \delta b_2}{(a_2^0)^2 + (b_2^0)^2 } \right]. \label{eq:delta_sigma}
\end{equation}

\subsection{Line depth} \label{sect:Ic}

The line depth $I_d$ is developed at first order as
\begin{align}
    I_d &= \frac{T}{2\sqrt{\pi} \sigma} \sqrt{a_1^2+b_1^2} \exp \left( \frac{\pi^2 \sigma^2}{T^2} \right) = \frac{T}{2\sqrt{\pi} \left(\sigma_0 + \delta\sigma\right)} \sqrt{(a_1^0)^2+(b_1^0)^2 + 2\delta a_1 a_1^0 + 2\delta b_1 b_1^0} \exp \left( \frac{\pi^2 (\sigma_0^2 + 2\sigma_0 \delta\sigma)}{T^2} \right), \\
    &= \frac{T}{2\sqrt{\pi} \sigma_0} \left[ \left(1 - \frac{\delta\sigma}{\sigma_0} \right) \sqrt{(a_1^0)^2+(b_1^0)^2} \left(1 + \frac{\delta a_1 a_1^0+\delta b_1 b_1^0}{(a_1^0)^2+(b_1^0)^2} \right) \exp \left( \frac{\pi^2 \sigma_0^2}{T^2} \right) \left(1 + \frac{2\pi^2}{T^2} \sigma_0 \delta\sigma \right) \right] = I_d^0 + \delta I_d,
\end{align}
where
\begin{equation}
    \delta I_d = I_d^0 \left[- \frac{\delta\sigma}{\sigma_0} +  \frac{\delta a_1 a_1^0+\delta b_1 b_1^0}{(a_1^0)^2+(b_1^0)^2} + \frac{2\pi^2\sigma_0 \delta\sigma}{T^2} \right],
\end{equation}
and $\delta\sigma$ is given by Eq.~\eqref{eq:delta_sigma}.

\subsection{Continuum intensity}

Using the previously derived expressions for the perturbed line width $\delta \sigma$ and line depth $\delta I_d$, we can compute the perturbed continuum intensity
\begin{align}
I_{\rm HMI} &= \frac{1}{N} \sum_{j=1}^N \left[ I_j + I_d \,\textrm{exp} \left(- \frac{(\lambda_j - \lambda_{\rm HMI})^2}{\sigma^2} \right) \right]= \frac{1}{6} \sum_{j=1}^N \left[ I_j^0 + \delta I_j  + (I_d^0 + \delta I_d) \,\textrm{exp} \left(- \frac{(\lambda_j - \lambda_{\rm HMI})^2}{\sigma_0^2 + 2 \sigma_0 \delta \sigma} \right) \right], \\
    &= \frac{1}{N} \sum_{j=1}^N \left[ I_j^0 + \delta I_j  + (I_d^0 + \delta I_d) \,\textrm{exp} \left(- \frac{(\lambda_j - \lambda_{\rm HMI})^2}{\sigma_0^2} \right) \left( 1 + 2\frac{(\lambda_j - \lambda_{\rm HMI})^2}{\sigma_0^2} \frac{\delta\sigma}{\sigma_0}  \right) \right], \\
    &= I_{\rm HMI}^0 + \frac{1}{N} \sum_{j=1}^N \left[ \delta I_j  +  \delta I_d \, \textrm{exp} \left(- \frac{(\lambda_j - \lambda_{\rm HMI})^2}{\sigma_0^2} \right)  \right] + I_d^0 \, \frac{\delta\sigma}{\sigma_0} \,\frac{2}{N} \sum_{j=1}^N \textrm{exp} \left(- \frac{(\lambda_j - \lambda_{\rm HMI})^2}{\sigma_0^2} \right) \frac{(\lambda_j - \lambda_{\rm HMI})^2}{\sigma_0^2} = I_{\rm HMI}^0 + \delta I_{\rm HMI},
\end{align}
where
\begin{equation}
    \delta I_{\rm HMI} = \frac{1}{N} \sum_{j=1}^N \left[\delta\mathcal{I}_j + \exp\left(- \frac{(\lambda_j - \lambda_{\rm HMI})^2}{\sigma_0^2} \right) \left( \delta I_d + 2 I_d^0 \frac{\delta\sigma}{\sigma_0} \frac{(\lambda_j - \lambda_{\rm HMI})^2}{\sigma_0^2}  \right)  \right].
\end{equation}

\section{Rewriting the thermodynamical contributions} \label{sect:thermodynamic}

We have shown in \citetalias{Kostogryz2021} that
\begin{equation}
    \delta I_{\tau,\alpha} = \int_0^{\tau^{\rm max}} S_\lambda^0 \textrm{e}^{-\tau_\lambda^0/\mu_0} \left( \frac{\delta \alpha_\lambda}{\alpha_\lambda^0} - \int_0^{\tau_\lambda^0} \frac{\delta \alpha_\lambda}{\alpha_\lambda^0} \frac{d\tau'_\lambda}{\mu_0} \right) \frac{d\tau_\lambda^0}{\mu_0}.
\end{equation}
Here, we give an equivalent form in order to make appear the background intensity $I_\lambda^0$. The second term can be integrated by part
\begin{align}
  \int_0^{\tau^{\rm max}} S_\lambda^0 \textrm{e}^{-\tau_\lambda^0/\mu_0}   \int_0^{\tau_\lambda^0} \frac{\delta \alpha_\lambda}{\alpha_\lambda^0} \frac{d\tau_\lambda'}{\mu_0} \frac{d\tau_\lambda^0}{\mu_0} &= - \int_0^{\tau^{\rm max}} \left[ \int_0^{\tau_\lambda^0} S_\lambda^0 \textrm{e}^{-\tau'_\lambda/\mu_0}  \frac{d\tau_\lambda'}{\mu_0} \right]  \frac{\delta \alpha_\lambda}{\alpha_\lambda^0}  \frac{d\tau_\lambda^0}{\mu_0} + \int_0^{\tau^{\rm max}}  S_\lambda^0 \textrm{e}^{-\tau_\lambda^0/\mu_0}  \frac{d\tau_\lambda^0}{\mu_0} \int_0^{\tau^{\rm max}} \frac{\delta \alpha_\lambda}{\alpha_\lambda^0} \frac{d\tau_\lambda^0}{\mu_0}, \\
  &= - \int_0^{\tau^{\rm max}} \left[ I_\lambda^0(0,\mu_0) - \int_{\tau_\lambda^0}^{\tau^{\rm max}} S_\lambda^0 \textrm{e}^{-\tau'_\lambda/\mu_0}  \frac{d\tau_\lambda'}{\mu_0} \right]  \frac{\delta \alpha_\lambda}{\alpha_\lambda^0}  \frac{d\tau_\lambda^0}{\mu_0} + I_\lambda^0(0,\mu_0) \int_0^{\tau^{\rm max}} \frac{\delta \alpha_\lambda}{\alpha_\lambda^0} \frac{d\tau_\lambda^0}{\mu_0}, \\
  &= \textrm{e}^{-\tau_\lambda^0 / \mu_0} \int_0^{\tau^{\rm max}} I_\lambda^0(\tau_\lambda^0,\mu_0) \frac{\delta \alpha_\lambda}{\alpha_\lambda^0}  \frac{d\tau_\lambda^0}{\mu_0}.
\end{align}
Thus
\begin{equation}
    \delta I_{\tau,\alpha} = \int_0^{\tau^{\rm max}} \Bigl[ S_\lambda^0  - I_\lambda^0(\tau_\lambda^0,\mu_0) \Bigr] \textrm{e}^{-\tau_\lambda^0/\mu_0} \,  \frac{\delta \alpha_\lambda}{\alpha_\lambda^0} \frac{d\tau_\lambda^0}{\mu_0},
\end{equation}
a form already obtained by \citet{Zhugzhda1996}.

\end{document}